\documentclass[9pt,shortpaper,twoside,web]{ieeecolor}
\usepackage{generic}
\usepackage{cite}
\usepackage{amsmath,amssymb,amsfonts}
\usepackage{algorithmic}
\usepackage{graphicx}
\usepackage{textcomp}
\usepackage{epstopdf}
\usepackage{subfigure}
\newtheorem{theorem}{Theorem}
\newtheorem{lemma}{Lemma}
\newtheorem{remark}{Remark}
\newtheorem{assumption}{Assumption}

\begin{document}
\title{Decentralized Dynamic Event-triggered Output-feedback Control of Stochastic Non-triangular Interconnected Systems with Unknown Time-varying Sensor Sensitivity}
\author{Libei~Sun,~Yongduan~Song,~\IEEEmembership{Fellow,~IEEE},~Maolong~Lv
\thanks{This work was supported in part by the National Key Research and Development Program of China under Grant 2022YFB4701400/4701401, in part by the CAA-Huawei MindSpore pen Fund, in part by the National Natural Science Foundation of China under Grant 61991400, Grant 61991403, Grant 62250710167, Grant 61860206008, Grant 61933012, and Grant 62273064),
{\color{black}{in part by National Natural Science Foundation of China under Grant 62303489 and Grant GKJJ24050502, and in part by Project for Science and Technology under Grant 2022-JCJQ-QT-018.}}}
\thanks{Libei Sun and Yongduan Song are with the Chongqing Key Laboratory of Intelligent Unmanned Systems, School of Automation, Chongqing University, Chongqing 400044, China (e-mail: lbsun@cqu.edu.cn; ydsong@cqu.edu.cn.  Corresponding author: Yongduan Song).}
\thanks{Maolong Lv is with the College of Air Traffic Control and Navigation, Air Force Engineering University, Xi'an, China 710051 (e-mail: maolonglv@163.com).}}
\maketitle

\begin{abstract}
This study addresses the intricate challenge of decentralized output-feedback control for stochastic non-triangular nonlinear interconnected systems with unknown time-varying sensor sensitivity in a dynamic event-triggered context. The presence of stochastic disturbances, non-triangular structural uncertainties, and evolving sensor sensitivity distinguishes this problem of global asymptotic stability from conventional event-triggered control scenarios. Existing event-triggered control approaches with static event conditions encounter difficulties in simultaneously ensuring zero tracking/stabilization error and preventing the occurrence of Zeno behavior. In this work, we develop a novel solution to address this complex issue. Firstly, we establish a linear relationship between the state vector of each interconnected subsystem and two error vectors through a unique coordinate transformation. This transformation effectively handles the complexities introduced by non-triangular structural uncertainties. Secondly, we introduce a decentralized dynamic event-triggered output-feedback control strategy, which involves a state observer and a decentralized output-feedback controller. Unlike conventional event-triggered control methods with static event conditions, this strategy formulates a modified clock-based dynamic triggering mechanism by introducing an auxiliary variable that evolves based on predicted plant state values, while utilizing a clock variable to guarantee the existence of a positive lower bound on inter-execution times. Rigorous Lyapunov analysis confirms the global asymptotic stability in probability of the closed-loop system, with the states and the output of each local subsystem converging to the equilibrium at the origin in probability.  Additionally, the existence of a minimal dwell-time between triggering instants is guaranteed. Finally, we substantiate the advantages and efficiency of the proposed algorithm through comprehensive numerical simulations.
\end{abstract}
\IEEEpeerreviewmaketitle

\section{Introduction}
\allowdisplaybreaks
In recent years, stochastic nonlinear systems have become a focal point of attention, given their significant roles in addressing various physical and engineering challenges. These systems are crucial when dealing with dynamic scenarios where random properties play a key role, such as random inputs arising from sensor noises, disturbances, and parameter changes \cite{fan2012decentralized}. The conventional approach of continuous-time state feedback control demands the ongoing monitoring of process states by the controller, leading to real-time decision-making. Consequently, this method incurs prohibitively high communication costs, making it impractical for many applications. To mitigate the communication burden, event-triggered control has emerged as an appealing alternative that enables the efficient transmission of measured signals based on predetermined triggering conditions \cite{tabuada2007event}, offering a practical  trade-off between achieving desired control performance and enhancing resource efficiency.

While a plethora of event-triggered control results exist in the current literature for deterministic nonlinear systems, covering  lower-order systems \cite{wang2020adaptive}, norm-form systems \cite{wang2021adaptive}, and strict-feedback systems \cite{sun_2022_distributed,xing2023dynamic}, there remains a notable dearth of research focused on  event-triggered control for stochastic nonlinear systems. This research gap is particularly critical in scenarios where the consideration of stochastic disturbances inherent in dynamic systems is essential. {\color{black}{For stochastic nonlinear systems, the direct application of the existing event-triggering mechanisms utilized in deterministic nonlinear systems  \cite{wang2020adaptive,wang2021adaptive,sun_2022_distributed,xing2023dynamic} may result in the occurrence of Zeno behavior. This is because the states of stochastic nonlinear systems can exceed any bound in an arbitrarily small amount of time, potentially resulting in the satisfaction of the triggered conditions within an arbitrarily small interval, i.e., a minimum positive inter-execution time may not be guaranteed.}} Addressing this challenge has prompted several efforts in recent years, including the exploration of time-regularization approach \cite{tang2022event} and two-stage method \cite{li2019event}. However, these approaches are only applicable to single systems with triangular structural uncertainties. 
%

The precision of measurement output hinges on the sensor sensitivity, a critical factor influencing feedback control design \cite{kolovsky2013nonlinear}. In practical engineering systems, like circuits, electrical devices, and mechanical systems, the sensor sensitivity in the system output is not consistently constant \cite{carr1993sensors}. To address the challenge of unknown time-varying sensor sensitivity, various output-feedback stabilization control schemes have been proposed in recent years.
For instance, a dual-domination approach was introduced in \cite{chen2017global} to tackle the global output-feedback stabilization problem for nonlinear systems with unknown measurement sensitivity. The work presented in \cite{li2020adaptive} suggested an adaptive output-feedback control scheme for stochastic nonlinear strict-feedback systems, considering both sensor uncertainty and unknown growth rates. Another study in \cite{li2020design} delved into the output-feedback control problem for stochastic feedforward systems with unknown measurement sensitivity. However, these results are limited to scenarios where plant state signals and the actuator signal are continuously transmitted over the network. This setup potentially imposes a communication burden on both the controller-to-actuator channel and the sensor-to-controller channel.

Despite the prevalence of stochastic disturbances and unknown time-varying sensor sensitivity, it is common in various applications for systems to exhibit non-triangular structural uncertainties, particularly in the dynamics of large-scale nonlinear uncertain interconnected systems \cite{zhou2008decentralized}. To address this issue, some attempts have been made in recent years. In \cite{cai2022decentralized}, a decentralized adaptive control scheme was developed for non-triangular nonlinear interconnected systems using backstepping techniques. Additionally, a robust adaptive backstepping output-feedback control algorithm was proposed in \cite{li2018robust} for non-triangular nonlinear systems with unmodeled dynamics, leveraging the stochastic small-gain theorem. However, both approaches presented in \cite{cai2022decentralized} and \cite{li2018robust} require the continuous transmission of plant state signals and the actuator signal over the network. More recently, the work in \cite{sun2023decentralized} investigated the decentralized event-triggered adaptive backstepping control problem in the presence of non-triangular structural uncertainties and time-varying parameters. Nevertheless, this solution is tailored for deterministic nonlinear systems. Furthermore,  the employed event-triggering condition is confined to a static form, ensuring only that the system tracking/stabilization error converges to a neighborhood of zero in the absence of Zeno behavior. To the best of our knowledge, within the framework of dynamic event-triggered control, there is no prior literature addressing the {\color{black}{decentralized output-feedback}} control problem for stochastic nonlinear interconnected systems while simultaneously considering both non-triangular structural uncertainties and unknown time-varying sensor sensitivity.

Motivated by the discussions above, we introduce a {\color{black}{decentralized dynamic event-triggered output-feedback control strategy}}  for stochastic non-triangular nonlinear interconnected systems, utilizing only the locally intermittent output signal. The development of this strategy addresses three main technical challenges: (i) \emph{Complex Subsystem Models}: Different from existing event-triggered control results \cite{wang2020adaptive,wang2021adaptive,sun_2022_distributed,sun2023decentralized,xing2023dynamic}, each subsystem model within the considered systems adopts a non-trivial nonlinear interconnected form with severe uncertainties, including stochastic disturbances, non-triangular structural uncertainties, and unknown time-varying sensor sensitivity.
(ii) \emph{Dynamic Event-triggered Communication and Limited Information}: In the context of dynamic event-triggered communication and unknown time-varying sensor sensitivity, only the locally intermittent output signal is accessible for controller design. This renders the widely employed backstepping design procedure \cite{deng1997stochastic,zhou2008decentralized,lv2022logic} impractical, as the repetitive differentiation of virtual control signals with respect to the triggering signals is no longer feasible due to the nature of event-triggering. 
(iii) \emph{Zeno Behavior Avoidance and Global Stability}: In existing event-triggered control results with static event conditions \cite{wang2021adaptive,sun2023decentralized,sun_2022_distributed}, the intentional incorporation of a positive constant is a common practice to exclude the occurrence of Zeno behavior. However, {\color{black}{these approaches}} only {\color{black}{ensure}} the convergence of tracking/stabilization error to a neighborhood around zero. {\color{black}{Besides, the existence of additive white noise disturbances in stochastic nonlinear systems presents a significant challenge to the straightforward adaptation of triggering mechanisms utilized in deterministic nonlinear systems, potentially resulting in the occurrence of Zeno behavior. 
Moreover, the diffusion terms in the considered systems depend on the states of the local subsystems and are not required to be bounded.  This poses a substantial challenge in control design and cannot be tackled by using the technique of its trivial generalization developed in \cite{zhou2008decentralized}.}}
Consequently, {\color{black}{existing approaches}} are not applicable within this research context. 
To address these challenges, in this study we offer a novel solution, a result not available priori in the literature. Firstly, we establish a crucial linear relationship between the state vector of each interconnected subsystem and two error vectors through the introduction of a novel coordinate transformation. This transformation effectively tackles challenges arising from non-triangular structural uncertainties.
Secondly, we propose a {\color{black}{decentralized dynamic event-triggered output-feedback control strategy,}} which intricately combines a state observer with a decentralized output-feedback controller.
{\color{black}{In this strategy, a modified clock-based dynamic triggering mechanism is formulated by introducing an auxiliary variable that evolves  according to predicted values of the plant states, while utilizing a clock variable to enforce a positive lower bound on inter-execution times, distinguishing it from conventional static event conditions.}}
We demonstrate that, the global asymptotic stability of the closed-loop system in probability is guaranteed, with the states and the output of each local subsystem converging to the equilibrium at the origin in probability. {\color{black}{Meanwhile, the designed modified clock-based dynamic triggering mechanism ensures the existence of a minimal dwell-time between any two triggering instants, thereby preventing the occurrence of Zeno behavior. This solution represents a significant advancement in addressing the challenges posed by stochastic non-triangular nonlinear interconnected systems with unknown time-varying sensor sensitivity in a dynamic event-triggered context.}}

\section{Problem Formulation and Preliminaries}
Consider a class of stochastic nonlinear systems composed of $N$ interconnected subsystems, with the $i$th subsystem modeled as:
\begin{align}
d{x}_{i, k}=\,&\left(x_{i, k+1}+\sum_{j=1}^{N} f_{i j, k}\left(x_{j}, u_{j}, t\right)\right)dt+\phi_{i,k}^{T}
\left(\check{x}_{i,k}\right)dw_i,&\nonumber\\
d{x}_{i, n_{i}}=\,&\left(u_{i}+\sum_{j=1}^{N} f_{i j, n_{i}}\left(x_{j}, u_{j}, t\right)\right)dt
+\phi_{i,n_i}^{T}\left({x}_{i}\right)dw_i,&\nonumber\\
y_{i}=\,&\theta_{i}(t)x_{i, 1},\, i=1,\cdots,N, & \label{eq:1}
\end{align}
for $k=1,\cdots, n_{i}-1$, where $x_i=[x_{i,1},\cdots,x_{i,n_i}]^{T}\in\mathcal{R}^{n_i}$, $u_i \in \mathcal{R}$, and $y_i\in \mathcal{R}$ are the state vector, the control input, and the output of the $i$th subsystem, respectively, the nonlinearity $\phi_{i,k}\left(\check{x}_{i,k}\right) {\color{black}{\in\mathcal{R}^{p}}}$ is a continuous function, with $\check{x}_{i,k}=[x_{i,1},\cdots,
x_{i,k}]^{T}\in\mathcal{R}^{k}$ and $\phi_{i,k}\left(0\right)=0$, $f_{i j, k}\left(x_{j}, u_{j}, t\right)\in \mathcal{R}$ denotes the nonlinear coupling interaction from the $j$th subsystem for $j \neq i$, or the modeling error of the $i$th subsystem for $j=i$, $\theta_i(t)$ represents the unknown time-varying sensor sensitivity, and $\omega_i$ {\color{black}{is an $p$-dimensional}} standard Wiener process.

The aim of this study is to develop {\color{black}{a decentralized dynamic event-triggered output-feedback control}} for the stochastic non-triangular nonlinear interconnected system (\ref{eq:1}) with unknown time-varying sensor sensitivity, utilizing only the locally intermittent output signal. The key objectives are as follows: i) ensuring the global asymptotic stability in probability of the closed-loop system, {\color{black}{with the states and the output of each local subsystem  converging to the equilibrium at the origin in probability}}; and ii) preventing the occurrence of Zeno behavior.

To realize these objectives, we need the following assumptions.
\begin{assumption}\label{assumption1}
The unknown nonlinear function $f_{i j, k}\left(x_{j}, u_{j}, t\right)$ satisfies:
\begin{align}
&\left|f_{i j, k}\left(x_{j}, u_{j}, t\right)\right| \leq \hbar_{i j, k}\left\|x_{j}\right\| ,k=1, \cdots, n_{i},& \label{eq:2}
\end{align}
for $i,j=1,\cdots, N$, where $\hbar_{ij,k}\ge{0}$ is the coupling gain that denotes the strength of the nonlinear coupling interaction, and there exists a known positive constant ${\bar{\hbar}_{ij,k}}$ such that $\hbar_{ij,k}\le {\bar{\hbar}_{ij,k}}$.
\end{assumption}

\begin{assumption}\label{assumption2}
There exists a positive constant $\ell_{i,k}$ with  a known upper bound $\bar{\ell}_{i,k}$, such that the following inequality holds for each $i=1,\cdots,N$, $k=1,\cdots,n_i$:
\begin{align}
\left|\phi_{i,k}\left(\check{x}_{i,k}\right)\right|
\leq \ell_{i,k}\left(\left|x_{i,1}\right|+\cdots
+\left|x_{i,k}\right|\right).& \label{eq:3}
\end{align}
\end{assumption}

\begin{assumption}\label{assumption3}
The sensor sensitivity $\theta_i(t)$ is an unknown time-varying function that adheres to $\theta_i(t)\in\left[1-\bar{\theta}_{i},
1+\bar{\theta}_{i}\right]$, 
where $0<\bar{\theta}_{i}<1$ is the allowable sensitivity error. 
\end{assumption}
\begin{lemma}\label{lemma2} (see \cite{chen2017global})
For $p \in[1, \infty)$ and any $x_{i,k} \in \mathcal{R}$, $i=1, \cdots, N$, $k=1,\cdots,n_i$, the following inequality holds:
\begin{align}
\left(\left|x_{i,1}\right|+\cdots
{\rm{+}}\left|x_{i,n_i}\right|\right)^p \leq n_i^{p-1}\left(\left|x_{i,1}\right|^p
+\cdots+\left|x_{i,n_i}\right|^p\right).&\label{eq:10}
\end{align}
\end{lemma}
\vspace{0.1cm}
\begin{remark}\label{remark1}
The common practice for dealing with the nonlinear coupling interaction $f_{ij,k}\left(x_{j}, u_{j}, t\right)$ entails the involvement of a condition as shown in Assumption~\ref{assumption1}  \cite{zhou2008decentralized}. It should be pointed out that, despite the utilizing of Assumption~\ref{assumption1}, a significant challenge persists.
This is because the nonlinear function $f_{ij,k}\left(x_{j}, u_{j}, t\right)$ in each local subsystem depends on the whole state vector $x_j=[x_{j,1},\cdots,x_{j,n_i}]^T$  $(j=1,\cdots,N)$, rather than the partial states vector $\check{x}_{j,k}=[x_{j,1},
\cdots,x_{j,k}]^T$ (as indicated in (\ref{eq:2})). Consequently, $f_{ij,k}\left(x_{j}, u_{j},t\right)$ does not adhere to a triangular structure of plant states. In this case, if the traditional backstepping control design procedure is employed, the virtual control functions $\alpha_{j,k}\,(k=1,\cdots,n_i)$ would be the function of whole state vector $x_j$, which would lead to the so-called algebraic loop problem and thus the traditional backstepping control design procedure is inapplicable. The existence of non-triangular structural uncertainties literally renders it rather difficult to address the stabilization problem of nonlinear interconnected systems \cite{cai2022decentralized,sun2023decentralized}.  Achieving global asymptotic stability for such systems becomes even more complicated when simultaneously considering both stochastic disturbances and unknown time-varying sensor sensitivity within the dynamic event-triggered control framework. The condition imposed on $\phi_{i,k}(\check{x}_{i,k})$ in Assumption~\ref{assumption2} is commonly employed in addressing the output-feedback control issue of stochastic nonlinear systems \cite{chen2017global,Deng1999}. As indicated in \cite{chen2017global,li2020adaptive}, Assumption~\ref{assumption3} is crucial to guarantee the stability of system (\ref{eq:1}).
\end{remark}

\begin{remark}\label{remark2}
This endeavor involves more than a straightforward extension from continuous output-feedback control \cite{chen2017global} to intermittent output-feedback control. Even in the absence of dynamic event-triggered communication, the presented results are novel and more intricate compared to the current body of knowledge \cite{chen2017global}. {\color{black}{Two significant technical challenges emerge in control design and stability analysis.}}
First, it is non-trivial to handle the impact of interactions in the design and analysis of local controllers which utilize only the locally available output signal for feedback, especially when these interactions lack a triangular structure. Consequently, the approach in \cite{chen2017global}, tailored for {\color{black}{single systems with triangular structural uncertainties,}} is inapplicable to the scenario considered in this study.
{\color{black}{Secondly, due to the existence of additive white noise disturbances, the structure and the fundamental theory for stochastic nonlinear systems are much more involved and difficult than that in the deterministic case \cite{chen2017global}.}}
\end{remark}

\section{State Observer Design}
In the presence of unknown time-varying sensor sensitivity, the states of the system are not directly available for control design. To address this challenge, we devise the following state observer to estimate the unmeasurable states of each local subsystem:
\begin{equation}\label{eq:11}
\begin{cases}
\dot{\hat{x}}_{i,1}=\hat{x}_{i,2}-L_{i,1}a_{i,1} \hat{x}_{i,1}, \\
\dot{\hat{x}}_{i,2}=\hat{x}_{i,3}-L_{i,1}^2a_{i,2} \hat{x}_{i,1}, \\
\quad \vdots\\
\dot{\hat{x}}_{i,n_i}=u_i-L_{i,1}^{n_i} a_{i,n_i} \hat{x}_{i,1},
\end{cases}
\end{equation}
where $L_{i,1}\ge{1}$ is {\color{black}{a constant gain to be specified designed later,}} and $a_{i,k}$ represents the positive coefficient of the Hurwitz polynomial $h_1(s) = s^{n_i} + a_{i,1}s^{n_i-1} + \cdots + a_{i,n_i-1}s + a_{i,n_i}$, $k = 1,\cdots,n_i$.

With the state observer (\ref{eq:11}), we introduce the coordinate transformation as follows
\begin{align}
{z_{i,1}} =&\, x_{i,1}, & \label{eq:12}\\
{z_{i,k}} =&\,\frac{1}{\left(L_{i,1}L_{i,2}\right)
^{k-1}} \hat{x}_{i,k},\,k = 2, \cdots ,n_i, & \label{eq:13}\\
{e_{i,k}} =&\,\frac{1}{L_{i,1}^{k-1}} \left({x_{i,k}-\hat{x}_{i,k}}\right),\,k = 1, \cdots ,n_i, & \label{eq:14}
\end{align}
where $L_{i,2}\ge{1}$ is {\color{black}{a constant gain to be designed later,}}
{\color{black}{with $e_i{\rm{=}}[e_{i,1},\cdots,e_{i,n_i}]^T$ and $z_i{\rm{=}}[z_{i,1},\cdots,z_{i,n_i}]^T$.}} From (\ref{eq:1}), (\ref{eq:11}), and (\ref{eq:14}), and using ${\rm{It\hat{o}}}$'s differentiation rule, one obtains
\begin{align}
d{e_{i,1}}{\rm{=}} &\left({L_{i,1}}{e_{i,2}}+ L_{i,1} a_{i,1} \hat{x}_{i,1}+\sum_{j=1}^{N} f_{i j, 1}\right)dt+\phi_{i,1}^{T}({x}_{i,1})
dw_i, &\nonumber\\
d{e_{i,k}}{\rm{=}} & \left(L_{i,1}e_{i,{k+1}}{\rm{+}}L_{i,1} a_{i,k}\hat{x}_{i,1}{\rm{+}}\sum_{j=1}^{N} \frac{f_{i j, k}}{L_{i,1}^{k-1}}\right)dt{\rm{+}}
\phi_{i,k}^{T}(\check{x}_{i,k})dw_i,
&\nonumber\\
d{e_{i,n_i}}{\rm{=}} &\left(L_{i,1} a_{i,n_i}\hat{x}_{i,1}+\sum_{j=1}^{N}\frac{f_{i j,n_i}}{L_{i,1}^{n_i-1}}\right)dt
+\phi_{i,n_i}^{T}\left({x}_{i}\right)dw_i, & \label{eq:15}
\end{align}
for $k=2,\cdots,n_i-1$. Then it follows that
\begin{align}
de_{i}=(L_{i,1}{A_e}{e_i}+L_{i,1}G_ex_{i,1}+F_e)dt+K_edw_i,
& \label{eq:16}
\end{align}
with
$$
A_e=\left[\begin{array}{cccc}
-a_{i,1} & 1 & \cdots & 0 \\
\vdots & \vdots & \ddots & \vdots \\
-a_{i,n_i-1} & 0 & \cdots & 1 \\
-a_{i,n_i} & 0 & \cdots & 0
\end{array}\right],
G_e=\left[\begin{array}{c}
a_{i,1} \\
\vdots \\
a_{i,n_i}
\end{array}\right],$$
$$
F_{e}=\left[\begin{array}{c}
\sum_{j=1}^{N} f_{i j, 1} \\
\sum_{j=1}^{N} \frac{1}{L_{i,1}}f_{i j, 2} \\
\vdots \\
\sum_{j=1}^{N} \frac{1}{L_{i,1}^{n_i-1}}f_{i j, n_i}
\end{array}\right],
K_e=\left[\begin{array}{c}
\phi_{i,1}^{T}
\left({x}_{i,1}\right) \\
\frac{1}{L_{i,1}}\phi_{i,2}^{T}
\left(\check{x}_{i,2}\right) \\
\vdots \\
\frac{1}{L_{i,1}^{n_i-1}}\phi_{i,n_i}^{T}
\left({x}_{i}\right)
\end{array}\right].$$
Since $A_e$ is a Hurwitz matrix, there exists a positive-definite matrix $P_i\in \mathcal{R}^{n_i\times n_i}$ such that $A_e^TP_i+P_iA_e=-I_{n_i}$, where $I_{n_i}\in \mathcal{R}^{n_i\times n_i}$ is an identity matrix. From the definition of $G_e$, it is seen that $\left\|G_e\right\|=({\sum_{k=1}^{n_i} a_{i,k}^2})^{1/2} \triangleq {r_e}$, where $r_e$ is some positive constant.

For the later development, we establish two useful lemmas that play an important role in establishing the main results of this study.
\begin{lemma}\label{lemma3}
The state vector $x_{i}$ obeys a linear relationship with two error vectors $z_{i}$ and $e_{i}$ as follows
\begin{align}
\left\|x_{i}\right\|^2 \leq \epsilon_{x_i}
\left\|z_{i}\right\|^2+{\color{black}{\sigma_{x_i}}}
\left\|e_{i}\right\|^2,i=1,\cdots,N, &\label{eq:17}
\end{align}
where $\epsilon_{x_i}$ and $\sigma_{x_i}$ are some positive constants that rely on {\color{black}{the constant gains $L_{i,1}$ and $L_{i,2}$.}}
\end{lemma}

{\emph{Proof}}. See Appendix I.

\begin{lemma}\label{lemma4}
The matrices $F_e$ and $K_e$ satisfy the following inequalities:
\begin{align}
\left\|F_e\right\|^2
\le & \sum_{j=1}^{N}\sum_{k=1}^{n_i}
\frac{1}{L_{i,1}^{2(k-1)}}{\color{black}{\hbar_{ij,k}^2}}
\left(\epsilon_{x_j}\left\|z_{j}\right\|^2
{\rm{+}}\sigma_{x_j}\left\|e_{j}\right\|^2\right),
&\label{eq:18}\\
\left\|K_e\right\|^2\le & \, \sum_{k=1}^{n_i}\frac{1}
{L_{i,1}^{2(k-1)}}k^2\ell_{i,k}^2 \left( \epsilon_{x_i}\left\|z_{i}\right\|^2+\sigma_{x_i}
\left\|e_{i}\right\|^2\right).&\label{eq:19}
\end{align}
\end{lemma}

{\emph{Proof}}. See Appendix II.

Consider a Lyapunov function candidate $V_{e_i}=e_i^TP_ie_i$. {\color{black}{By using the ${\rm{It\hat{o}}}$'s differentiation rule and Young's inequality, the infinitesimal generator of $V_{e_i}$ can be derived from (\ref{eq:16}) that}}
\begin{align}
\mathcal{L} V_{e_i}
\le&-\frac{1}{2}L_{i,1}
\left\|e_i\right\|^2+\frac{1}{2}\left\|P_i\right\|^2
\left\|e_i\right\|^2+2L_{i,1}{r_e^2}\left\|P_i\right\|^2
\left\|z_{i}\right\|^2 &\nonumber\\
&+\sum_{j=1}^{N}\sum_{k=1}^{n_i}
\frac{2}{L_{i,1}^{2(k-1)}}\hbar_{ij,k}^2\left(
\epsilon_{x_j}\left\|z_{j}\right\|^2 +\sigma_{x_j}\left\|e_{j}\right\|^2\right)\nonumber\\
&+\sum_{k=1}^{n_i}
\frac{k^2}{L_{i,1}^{2(k-1)}}\ell_{i,k}^2
\|P_i\|(\epsilon_{x_i}\|z_{i}\|^2{\rm{+}}
\sigma_{x_i}\|e_{i}\|^2).&\label{eq:25}
\end{align}
Define a Lyapunov function candidate for the overall closed-loop system as $V_{e}=\sum_{j=1}^{N}V_{e_j}$. With the aid of (\ref{eq:25}), we obtain
\begin{align}
\mathcal{L}V_{e}\le-\sum_{j=1}^{N}\sigma_{e_j}
\left\|e_{j}\right\|^2+\sum_{j=1}^{N}\sigma_{z_j}
\left\|z_{j}\right\|^2,&\label{eq:27}
\end{align}
where $\sigma_{e_j}=\frac{1}{2}L_{j,1}-\frac{1}{2}
\left\|P_j\right\|^2-2\sum_{i=1}^{N}
{\color{black}{\sum_{k=1}^{n_i}
\frac{1}{L_{j,1}^{2(k-1)}}\hbar_{ij,k}^2\sigma_{x_j}}}
-{\color{black}{\sum_{k=1}^{n_i} }}
\frac{1}{L_{j,1}^{2(k-1)}}k^2\ell_{j,k}^2\sigma_{x_j}\left\|P_j\right\|$ and $\sigma_{z_j}=2L_{j,1}{r_e^2}\left\|P_j\right\|^2
+\sum_{i=1}^{N}\sum_{k=1}^{n_i}\frac{2}{L_{j,1}^{2(k-1)}}\hbar_{ij,k}^2\epsilon_{x_j}
+{\color{black}{\sum_{k=1}^{n_i}}} \frac{1}{L_{j,1}^{2(k-1)}}k^2\ell_{j,k}^2 \epsilon_{x_j}\left\|P_j\right\|$.

\begin{remark}\label{remark3}
The sensor sensitivity plays a crucial role in the control design of feedback control systems, as it directly influence the accuracy of the measurement output. In various practical engineering systems, such as circuits, electrical devices, and mechanical systems, the sensor sensitivity associated with the system output undergoes dynamic variations and does not remain constant. However, the current state-of-the-art on output-feedback control for stochastic nonlinear systems lacks consideration for the impact of sensor sensitivity on the system output \cite{fan2012decentralized,li2018robust}.
To address this gap, we introduce a state observer to estimate unmeasurable states for each local subsystem. The constructed observer, as expressed in (\ref{eq:11}), is more general and essentially different from related results \cite{Deng1999,xing2018event,zhou2008decentralized,fan2012decentralized}. This distinction arises for two main reasons.
Firstly, the design of the state observer (\ref{eq:11}) is independent of information related to non-triangular structural uncertainties $f_{i j, k}\left(x_{j}, u_{j}, t\right)$ and nonlinearities $\phi_{i,k}\left(\check{x}_{i,k}\right)$. {\color{black}{This renders it simple and straightforward to implement, in contrast to existing observer-based approaches that rely on the information of system models  \cite{xing2018event,zhou2008decentralized}}}. 
Secondly, unlike \cite{Deng1999,fan2012decentralized}, the system (\ref{eq:1}) includes sensor sensitivity $\theta_i(t)$, {\color{black}{which is both unknown and time-varying.}} The presence of such sensor sensitivity prevents the direct utilization of $x_{i,1}$ in establishing both the state observer and control laws. Therefore, the constructed state observer (\ref{eq:11}) demonstrates robustness against measurement and structure uncertainties inherent in system (\ref{eq:1}).
\end{remark}

\begin{remark}\label{remark4}
{\color{black}{An observer error dynamic system operating in a fully open-loop configuration, where both the control input $u_i$ and the output $y_i$ are absent, is inherently unstable. Conversely, when the observer error dynamic system includes the control input $u_i$ but excludes the output $y_i$, it is termed a semi-closed loop. In this setup, if the input $u_i$ is appropriately designed within the co-designed system that includes both the observer error dynamic system and the control system, as we have done in this study, the resulting observer error dynamic system is ensured to be asymptotically stable for deterministic systems and asymptotically stable in probability for stochastic systems. This assertion has been authenticated by our detailed theoretical analysis and is well-supported by a comprehensive body of literature on observer design and analysis. Noteworthy references include the works of \cite{chen2017global} (Equation (4)), \cite{li2020adaptive} (Equations (11)-(13)), \cite{li2022prescribed} (Equations (7)-(9)), \cite{qian2002output} (Equation (2.1)).  These studies offer extensive evidence for the effectiveness of the observer design proposed in our study, thereby underpinning the validity of our approach.}}
\end{remark}

\section{Decentralized Continuous Output-feedback Control}
To better illustrate the control design approach employed in this work, let us initially consider the stochastic non-triangular nonlinear interconnected systems described by (\ref{eq:1}) under continuous output feedback. This serves as the foundation for the control scheme under  intermittent output feedback in next section. 

Building upon the state observer presented in (\ref{eq:11}), we introduce a decentralized {\color{black}{continuous output-feedback control scheme}} as follows
\begin{align}
u_{i}=&-\kappa_i\left(b_{i,n_i}
y_i+\frac{1}{L_{i,1}L_{i,2}}b_{i,n_i-1}\hat{x}_{i,2}+\cdots\right.& \nonumber\\
&\left. +\frac{b_{i,2}}{(L_{i,1}L_{i,2})^{n_i-2}}
\hat{x}_{i,n_i-1}+\frac{b_{i,1}}{(L_{i,1}L_{i,2})^{n_i-1}}
\hat{x}_{i,n_i} \right),&\label{eq:31}
\end{align}
where $\kappa_i={\left(L_{i,1}L_{i,2}\right)^{n_i}}$ and $b_{i,k}(k = 1,\cdots,n_i$) is the positive coefficient of the Hurwitz polynomial $h_2(s) = s^{n_i} + b_{i,1}s^{n_i-1} + \cdots + b_{i,n_i-1}s + b_{i,n_i}$.

Now we are ready to state the following theorem.
\begin{theorem}\label{theorem1}
Consider the stochastic non-triangular nonlinear interconnected system (\ref{eq:1}) under Assumptions~\ref{assumption1}-\ref{assumption3}.  If the state observer (\ref{eq:11}) and the decentralized {\color{black}{continuous output-feedback controller}} (\ref{eq:31}) are employed, the following conclusions hold:
\begin{itemize}
\item [i)]{The global asymptotic stability of the closed-loop system in probability is guaranteed.}
\item [ii)]{{\color{black}{The states and the output of each local subsystem are ensured to converge to the equilibrium at the origin in probability.}}}
\end{itemize}
\end{theorem}

${\emph{Proof}}$.
{\color{black}{The proof involves two parts: stability analysis and determination of the constant gains.

Part I. Stability analysis.}}
Utilizing (\ref{eq:12}), (\ref{eq:13}), and (\ref{eq:14}), one can derive
\begin{align}
x_{i,1}=&z_{i,1},&\label{eq:33}\\
x_{i,k}=&{L_{i,1}^{k-1}}{e_{i,k}}+ {\left(L_{i,1}L_{i,2}\right)^{k-1}}{z_{i,k}},& \label{eq:34}
\end{align}
for $k = 2, \cdots ,n_i$. According to the ${\rm{It\hat{o}}}$'s differentiation rule, and utilizing (\ref{eq:1}), (\ref{eq:11}), (\ref{eq:12}), (\ref{eq:13}), and (\ref{eq:34}), we have 
\begin{align}
d{z_{i,1}}= &\left({L_{i,1}}{e_{i,2}}{\rm{+}} L_{i,1}L_{i,2}{z_{i,2}}+{\color{black}{\sum_{j=1}^{N} f_{i j, 1}}}\right)
dt+\phi_{i,1}^{T}\left({x}_{i,1}\right)dw_i,&\nonumber\\
d{z_{i,k}}=&\left(L_{i,1}L_{i,2}
{\color{black}{z_{i,k+1} }} +\frac{L_{i,1}}{L_{i,2}^{k-1}}
a_{i,k}(e_{i,1}-x_{i,1})\right)dt, &\nonumber\\
d{z_{i,n_i}}=& \left(\frac{1}{(L_{i,1}L_{i,2})
^{n_i-1}}u_i+\frac{L_{i,1}}{L_{i,2}^{n_i-1}} a_{i,n_i}(e_{i,1}{\rm{-}}x_{i,1}) \right)dt,& \label{eq:35}
\end{align}
for $k=2,\cdots,n_i-1$. Inserting (\ref{eq:31}) into (\ref{eq:35}) yields
\begin{align}
d{z_{i}}=&\left(L_{i,1}L_{i,2}{A_z}{z_i}+L_{i,1}L_{i,2}
{R}_z{b_{i,n_i}}\left(1-\theta_i\right)
z_{i,1}+F_z\right.& \nonumber\\
&\left.+\frac{L_{i,1}}{L_{i,2}}G_z
(e_{i,1}-z_{i,1})+L_{i,1}D_ze_{i,2}\right)dt
+K_zdw_i,& \label{eq:36}
\end{align}
with 
$$
A_z{\rm{=}}\left[\begin{array}{cccc}
0 & 1 & \cdots & 0 \\
\vdots & \vdots & \ddots & \vdots \\
0 & 0 & \cdots & 1 \\
-b_{i,n_i} & -b_{i,n_i-1} & \cdots & -b_{i,1}
\end{array}\right],
{R}_{z}{\rm{=}}\left[\begin{array}{c}
0 \\
0 \\
\vdots \\
1
\end{array}\right],
D_{z}{\rm{=}}\left[\begin{array}{c}
1 \\
0 \\
\vdots \\
0
\end{array}\right],
$$
$$
G_z{\rm{=}}\left[\begin{array}{c}
0 \\
a_{i,2} \\
\vdots \\
\frac{1}{L_{i,2}^{n_i-2}}a_{i,n_i}
\end{array}\right],
F_{z}{\rm{=}}\left[\begin{array}{c}
\sum_{j=1}^{N}f_{i j, 1} \\
0 \\
\vdots \\
0
\end{array}\right],
K_{z}=\left[\begin{array}{c}
{\color{black}{\phi_{i,1}^T}} \\
0 \\
\vdots \\
0
\end{array}\right].
$$
Since $A_z$ is a Hurwitz matrix, there is a positive definite matrix $Q_i$ such that $A_z^{T}Q_i +Q_i{\color{black}{A_z}}=-I_{n_i}$. By the definition of ${R}_z$, $D_z$, and $G_z$, {\color{black}{it is readily seen that}} $\|{R}_z\|=\|D_z\|=1$, and $\|G_z\|={(\sum_{k=2}^{n_i} a_{i,k}^2/{L_{i,2}^{2(k-2)}})}^{{1}/{2}} \triangleq {r_z}$, where $r_z$ is some positive constant.

Consider a Lyapunov function candidate $V_{z_i}=z_i^TQ_iz_i$. From (\ref{eq:36}) and applying the ${\rm{It\hat{o}}}$'s differentiation rule, $\mathcal{L} V_{z_i}$ can be derived as
\begin{align}
\mathcal{L} V_{z_i}=&2z_i^TQ_i\left(L_{i,1}
L_{i,2}{A_z}{z_i}+\frac{L_{i,1}}{L_{i,2}}G_z
\left(e_{i,1}{\rm{-}}z_{i,1}\right)
+L_{i,1}D_ze_{i,2}\right.& \nonumber\\
&\left. +L_{i,1}L_{i,2}{R}_z{b_{i,n_i}}
\left(1{\rm{-}}\theta_i\right)z_{i,1} +F_z \right) +Q_i\left\|K_z\right\|^2.&\label{eq:39}
\end{align}
{\color{black}{By recalling Lemma~\ref{lemma3} and employing Young's inequality leads to}}
\begin{align}
\mathcal{L} V_{z_i}\le &{\rm{-}}L_{i,1}L_{i,2}\left\|z_i\right\|^2+
2 L_{i,1}L_{i,2}{b_{i,n_i}}\left\|Q_i\right\|
\left|1-\theta_i\right|\left\|z_{i}\right\|^2
&\nonumber\\
&{\rm{+}}{\color{black}{\frac{2L_{i,1}^2}{L_{i,2}^2}
{r_z^2}\left\|z_i\right\|^2}}+
\frac{2L_{i,1}}{L_{i,2}}r_z \left\|Q_i\right\| \left\|z_i\right\|^2  +2\left\|Q_i\right\|^2
\left\|z_{i}\right\|^2&\nonumber\\
&+ \ell_{i,1}^2\|Q_i\|\|z_{i}\|^2
{\rm{+}}\sum_{j=1}^{N}\frac{1}{2}\hbar_{ij,1}^2
\left(\epsilon_{x_j}\left\|z_{j}\right\|^2
+\sigma_{x_j}\left\|e_{j}\right\|^2\right)
&\nonumber\\
&+ {\color{black}{2L_{i,1}}}\left\|z_i\right\|^2
+{\color{black}{\frac{1}{2}\|Q_i\|^2
\|e_i\|^2}}+\frac{1}{2}L_{i,1}{\color{black}{\|Q_i\|^2}} \|e_i\|^2.&\label{eq:48}
\end{align}
{\color{black}{Consider a Lyapunov function candidate for the entire interconnected system as $V_{z}=\sum_{j=1}^{N}V_{z_j}$. From (\ref{eq:48}), it can be deduced that}}
\begin{align}
\mathcal{L} V_{z}\le -\sum_{j=1}^{N}\rho_{z_j}
\left\|z_j\right\|^2+\sum_{j=1}^{N}\rho_{e_j}
\left\|e_j\right\|^2,&\label{eq:50}
\end{align}
where $\rho_{z_j}=L_{j,1}L_{j,2}-{\color{black}{\frac{2 L_{j,1}^2}{L_{j,2}^2}r_z^2}}-{\color{black}{2L_{j,1}}} -\ell_{j,1}^2\left\|Q_j\right\|-2\left\|Q_j\right\|^2
-\frac{2 L_{j,1}}{L_{j,2}}
r_z\left\|Q_j\right\|-\sum_{i=1}^{N}\frac{1}{2} \hbar_{ij,1}^2\epsilon_{x_j}-2L_{j,1}L_{j,2}
{b_{j,n_j}}\left\|Q_j\right\|\left|1-\theta_j\right|$  and $\rho_{e_j}=\frac{1}{2}L_{j,1} {\color{black}{\left\|Q_j\right\|^2}}+{\color{black}{\frac{1}{2} \left\|Q_j\right\|^2}}+\sum_{i=1}^{N}\frac{1}{2} \hbar_{ij,1}^2\sigma_{x_j}$.

At this point, we consider a Lyapunov function candidate $V=V_{e}+V_{z}$. According to  (\ref{eq:27}) and (\ref{eq:50}), $\mathcal{L}V$ can be upper bounded by
\begin{align}
\mathcal{L}V\le -\sum_{j=1}^{N}\xi_j\left\|e_{j}\right\|^2
-\sum_{j=1}^{N}\varsigma_j \left\|z_{j}\right\|^2,
& \label{eq:54}
\end{align}
where $\xi_{j}=\sigma_{e_j}
-\rho_{e_j}$, $\varsigma_j=\rho_{z_j}-\sigma_{z_j}$. 
\vspace{0.1cm}

Part II. Determination of the constant  gains $L_{j,1}$ and $L_{j,2}$. Let $\xi_{j}>0$, the following inequality holds
\begin{align}
\frac{1}{2}L_{j,1}- \frac{1}{2}L_{j,1} {\color{black}{\|Q_j\|^2}}>\hat{c}_1,
&\label{eq:R2}
\end{align}
where
$\hat{c}_1=\frac{1}{2}\left\|P_j\right\|^2
+2\sum_{i=1}^{N}{\color{black}{\bar{\hbar}_{ij,k}^2}}
\sigma_{x_j}+k^2{\color{black}{\bar{\ell}}}_{j,k}^2
\sigma_{x_j}\left\|P_j\right\|
+{\color{black}{\frac{1}{2}\left\|Q_j\right\|^2}}
+\sum_{i=1}^{N}\frac{1}{2}{\color{black}{\bar{\hbar}
_{ij,1}^2}}\sigma_{x_j}>0$ is a computable constant.
{\color{black}{This, combined with $L_{j,1}>1$, results in}}
\begin{align}
L_{j,1}=\max\left\{1,\frac{2\hat{c}_1}
{1-{\color{black}{\left\|Q_j\right\|^2}}}\right\}.
&\label{eq:R4}
\end{align}
Let $\varsigma_j>0$, there holds $(1-2{b_{j,n_j}}\left\|Q_j\right\|\left|1
-\theta_j\right|)L_{j,1}L_{j,2}-2({r_e^2}
\left\|P_j\right\|^2+1)L_{j,1}-2 r_z^2{\color{black}{\frac{ L_{j,1}^2}{L_{j,2}^2}}}-2 r_z\left\|Q_j\right\|\frac{ L_{j,1}}{L_{j,2}}>\hat{c}_2$,
where $\hat{c}_2=\bar{\ell}_{j,1}^2 \left\|Q_j\right\|
+2\left\|Q_j\right\|^2+\sum_{i=1}^{N}
\frac{1}{2}{\color{black}{\bar{\hbar}_{ij,1}^2}}
\epsilon_{x_j}+\sum_{i=1}^{N}{\color{black}{2\bar{\hbar}_{ij,k}^2}}
\epsilon_{x_j}+k^2{\color{black}{\bar{\ell}_{j,k}^2}}
\epsilon_{x_j}\left\|P_j\right\|>0$ is a computable constant. By choosing $\bar{\theta}_j$ as $\bar{\theta}_j<\theta_j^*=\frac{1}{2 b_{j,n_j}\|Q_j\|}$, where $\theta_j^*$ is the upper-bound of the allowable sensitivity error. Since $1-\bar{\theta}_j \leq \theta_j \leq 1+\bar{\theta}_j$ and $\bar{\theta}_j<\theta_j^*$, it has $1-2 b_{j,n_j}|1-\theta_j|\|Q_j\|\geq 1-2 b_{j,n_j} \bar{\theta}_j\|Q_j\|\triangleq \Lambda_j \in (0,1)$. Then, it is seen that $\varsigma_j>0$ is valid if
$L_{j,1}L_{j,2}-a_1\frac{L_{j,1}^2}{L_{j,2}^2}
-a_2L_{j,1}-a_3\frac{L_{j,1}}{L_{j,2}}
>\frac{1}{\Lambda_j}\hat{c}_2$,
where $a_1={\color{black}{\frac{2}{\Lambda_j} }} r_z^2>0$, $a_2={\color{black}{\frac{2}{\Lambda_j} }}({r_e^2}\left\|P_j
\right\|^2+1)>0$, and $a_3={\color{black}{\frac{ 2}{\Lambda_j} }}r_z\left\|Q_j\right\|>0$ are computable constants.
Furthermore, using $L_{j,2}>1$, $\varsigma_j>0$ holds true if
\begin{align}
L_{j,2}=\max\left\{1,{\kappa_1}{\color{black}{\hat{c}_2}}
+{\kappa_2}\right\},&\label{eq:AR8}
\end{align}
where ${\kappa_1}=\frac{1}{\Lambda_jL_{j,1}}>0$ and ${\kappa_2}=a_1L_{j,1}+a_2+a_3>0$ are some computable constants.

From the above analysis, it is seen that the constant gains $L_{j,1}$ and $L_{j,2}$ can be chosen properly to ensure $\xi_j>0$ and $\varsigma_j>0$ by following the design recipe provided in (\ref{eq:R4}) and (\ref{eq:AR8}), respectively.

Now we are ready to establish the validity of the results presented in Theorem~\ref{theorem1}.

i) From (\ref{eq:54}) {\color{black}{and using the design recipe of constant gains provided in (\ref{eq:R4}) and (\ref{eq:AR8}),}} it is readily seen that $\mathcal{L}V$ is negative definite.
{\color{black}{Then,}} by applying Theorem 2.1 in \cite{deng1997stochastic}, we can conclude that the closed-loop system is globally asymptotically stable in probability.

ii) By employing Dynkin's formula \cite{Hasminskii1980stochastic}, it is deduced from (\ref{eq:54}) that $E\left[V(t)\right]\le V(0)e^{-ct},\forall t\ge{0}$, where $c=\min\{\xi_{j}, \varsigma_j\}>0$. By the definition of $V$, it is seen that $e_{i,k}$ and $z_{i,k}$ ($k=1,\cdots,n_i$) are ensured to converge to the equilibrium at the origin in probability. Then, from (\ref{eq:33}) and (\ref{eq:34}), and using $y_i=\theta_ix_{i,1}$  (with $1-\bar{\theta}_{i}\le \theta_i\le 1+\bar{\theta}_{i}$), it is established that the states $x_{i,k}(k=1,\cdots,n)$ and the output $y_i$ of each local subsystem are ensured to converge to the equilibrium at the origin in probability. $\hfill\blacksquare$

\section{Decentralized Dynamic Event-triggered Output-feedback Control}
In this section, we extend the results to a more general case where both the actuator signal and the output signal are permitted to be triggered. Consequently, only the locally intermittent output signal is available for control design.  The decentralized dynamic {\color{black}{event-triggered output-feedback control scheme}} is constructed by replacing the continuous signals in the preceding scheme with the triggered ones. The stability analysis of this extended scheme, which is a key focus of this section, demands a delicate way to ensure the boundedness of triggering errors, {\color{black}{the prevention of the occurrence of Zeno behavior,}} and the global asymptotic stability of the closed-loop system.

\subsection{Design of A Modified Clock-based Dynamic Triggering Mechanism}
To ensure the global asymptotic stability of the stochastic non-triangular nonlinear interconnected system (\ref{eq:1}), while concurrently preventing the occurrence of Zeno behavior, {\color{black}{we introduce a modified clock-based dynamic triggering mechanism inspired partly by the ideas in \cite{xing2023dynamic,tang2022event}, which is formulated as}}
\begin{align}
{\bar{y}}_{i}\left( t \right){\rm{=}}&y_i({t_{y,l}^{i}}),
\forall t\in [t_{y,l}^{i},t_{y,l + 1}^{i}),
&\label{eq:M55}\\
t_{y,l+1}^i{\rm{=}} &\inf\{{t{\rm{>}} t_{y,l}^i
\vert \Xi_{y_i}{\rm{\le}} \rho_{y_i}(\varpi_{y_i}^2{\rm{-}}\alpha_{y_i}
\|z_i^{*}\|^2)}{\color{black}{\,\,\&\, \Pi_{y_i}\le{\underline{\Pi}_{y_i}}}}\},
&\label{eq:55}\\
{u_i}\left( t \right){\rm{=}}& {v_i}( {t_{u,l}^{i}}),
\forall t \in [t_{u,l}^{i},t_{u,l + 1}^{i}),
&\label{eq:M58}\\
t_{u,l+1}^i{\rm{=}} &\inf\{{t{\rm{>}}t_{u,l}^i
\vert\Xi_{u_i}{\rm{\le}}\rho_{u_i}(\varpi_{u_i}^2{\rm{-}}\alpha_{u_i}
\|z_i^{*}\|^2)\&\Pi_{u_i}\le{\underline{\Pi}_{u_i}}}\},
& \label{eq:58}
\end{align}
where $t_{y,l}^{i}$ and $t_{u,l}^{i}$ ($l=0,1,2,\cdots$) represent the $l$th event times at which the output signal and the actuator signal broadcast their information, respectively, $v_i$ is the controller to be designed, $u_i$ is the final controller, denoting the controller after triggering, and $\bar{y}_i$ is the output after triggering.
$\varpi_{y_i}=y_i-\bar{y}_i$ and $\varpi_{u_i}=v_i-u_i$ are triggering errors of the output signal and the actuator signal, respectively, {\color{black}{$\rho_{y_i}$, $\rho_{u_i}$, $\alpha_{y_i}$, $\alpha_{u_i}$, $\underline{\Pi}_{y_i}$, and $\underline{\Pi}_{u_i}$ are some positive design parameters, and $z_i^{*}=[0,z_{i,2},\cdots,z_{i,n_i}]^T$ is the coordinate transformation vector. It is noteworthy that we employ $z_i^{*}$ rather than $z_i$ in the dynamic triggering conditions (\ref{eq:55}) and (\ref{eq:58}), as the time-varying sensor sensitivity renders $z_{i,1}$ unavailable.}}
The dynamic auxiliary variables ${\Xi}_{y_i}$ and ${\Xi}_{u_i}$ are updated in the following manner
\begin{align}
\dot{\Xi}_{y_i}=&-\beta_{y_i}\Xi_{y_i}+\gamma_{y_i}
\left(\alpha_{y_i}\left\|z_i^{*}\right\|^2
-\varpi_{y_i}^2\right),\,\,\Xi_{y_i}(0)>0,
& \label{eq:59}\\
\dot{\Xi}_{u_i}=&-\beta_{u_i}\Xi_{u_i}+\gamma_{u_i}
\left(\alpha_{u_i}\left\|z_i^{*}\right\|^2
{\rm{-}}\varpi_{u_i}^2\right),\,\,\Xi_{u_i}(0)\,{\rm{>}}\,0,
& \label{eq:62}
\end{align}
where {\color{black}{$\beta_{y_i}$, $\beta_{u_i}$, $\gamma_{y_i}$, and $\gamma_{u_i}$ are some positive design parameters, which satisfy  ${\rho_{y_i}}>\frac{1- \gamma_{y_i}}{\beta_{y_i}}$ and ${\rho_{u_i}}>\frac{1- \gamma_{u_i}}{\beta_{u_i}}$.}}
From (\ref{eq:55}) and (\ref{eq:58}), it is seen that $\alpha_{y_i}\left\|z_i^{*}\right\|^2
-\varpi_{y_i}^2\ge -{\frac{\Xi_{y_i}}
{\rho_{y_i}}}$ in the interval $[t_{y,l}^i,t_{y,l + 1}^i)$, and $\alpha_{u_i}\left\|z_i^{*}
\right\|^2-\varpi_{u_i}^2\ge -{\frac{\Xi_{u_i}}{\rho_{u_i}}}$ in the interval $[t_{u,l}^i,t_{u,l + 1}^i)$. Then it can be derived from (\ref{eq:59}) and (\ref{eq:62}) that
$\dot{\Xi}_{y_i}\ge -\left(\beta_{y_i}+\frac{\gamma_{y_i}}
{\rho_{y_i}}\right)\Xi_{y_i}$ and $\dot{\Xi}_{u_i}\ge-\left(\beta_{u_i}+\frac{\gamma_{u_i}}
{\rho_{u_i}}\right)\Xi_{u_i}$.
This, combined with $\Xi_{y_i}(0)>0$ and $\Xi_{u_i}(0)>0$, results in
${\Xi}_{{y_i}}(t)>0$ and ${\Xi}_{{u_i}}(t)>0$.  {\color{black}{The dynamics of the clock variables $\Pi_{y_i}$ and $\Pi_{u_i}$ are given by
\begin{align}
	\dot{\Pi}_{y_i}=&- \delta_{y_i}  \left(\varpi_{y_i}^2+ \Pi_{y_i}^2\right),\,\,\,
	\Pi_{y_i}(0)=\bar{\Pi}_{y_i},& \label{eq:A62}\\
	\dot{\Pi}_{u_i}=&-\delta_{u_i}   \left(\varpi_{u_i}^2+ \Pi_{u_i}^2\right),\,\,
	\Pi_{u_i}(0)=\bar{\Pi}_{u_i},& \label{eq:B62}
\end{align}
where $\delta_{y_i}$, $\delta_{u_i}$, $\bar{\Pi}_{y_i}$, and $\bar{\Pi}_{u_i}$ are some positive design parameters, with $\bar{\Pi}_{y_i}>\underline{\Pi}_{y_i}$, $\bar{\Pi}_{u_i}>\underline{\Pi}_{u_i}$. The clock variables at each triggering instant are defined as $\Pi_{y_i}\left(t_{y,l}^{i}\right)=\bar{\Pi}_{y_i}$ and $\Pi_{u_i}\left(t_{u,l}^{i}\right)=\bar{\Pi}_{u_i}$, respectively.}} 

\subsection{Discussion on the Design Philosophy and Merits of the Proposed Triggering Mechanism}
{\color{black}{Firstly, we elucidate the underlying design philosophy behind the proposed modified clock-based dynamic triggering mechanism  (here we exemplify with the output triggering mechanism).		
		
The proposed output triggering condition (\ref{eq:55}) consists two parts, each playing a cricual role. In the first part, a dynamic auxiliary variable ${\Xi}_{y_i}$ that evolves based on predicted plant state values, is introduced to ensure zero stabilization error (as demonstrated in Section V-C). In the second part, a clock variable $\Pi_{y_i}$ is utilized to guarantee the existence of a positive lower bound on inter-execution times. The idea behind involving the clock variable is as follows:
Since $\Pi_{y_i}$ is initialized as the upper bound $\bar{\Pi}_{y_i}$ when the $l$th event is triggered at time instant $t_{y,l}^{i}$, according to the dynamics of ${\Pi}_{y_i}$ in (\ref{eq:A62}) and the second part of the triggering condition (\ref{eq:55}), $\Pi_{y_i}$ begins to decrease from $\bar{\Pi}_{y_i}$, and upon reaching the lower bound $\underline{\Pi}_{y_i}$, the second part of (\ref{eq:55}) is met. Should the first part of (\ref{eq:55}) also be met at this moment, $\Pi_{y_i}$ will be reset to $\bar{\Pi}_{y_i}$ again, as the $(l+1)$th event is triggered at time instant $t_{y,l+1}^{i}$. Otherwise,  the event will not be triggered.
According to (\ref{eq:A62}), a larger $\Pi_{y_i}$ (or $\varpi_{y_i}$) leads to a larger $\dot{\Pi}_{y_i}$, causing a faster reduction in $\Pi_{y_i}$. This indicates that the inter-event interval will not consistently increase with a larger $\bar{\Pi}_{y_i}$ (or  $\varpi_{y_i}$). Conversely, a smaller $\Pi_{y_i}$ (or $\varpi_{y_i}$) produces a smaller $\dot{\Pi}_{y_i}$, leading to a slower reduction in $\Pi_{y_i}$. This suggests that the inter-event interval will not consistently decrease with a smaller ${\Pi}_{y_i}$ (or $\varpi_{y_i}$).
Consequently, the proposed triggering mechanism allows ${\Pi}_{y_i}$ to decrease from $\bar{\Pi}_{y_i}$ to $\underline{\Pi}_{y_i}$ within a finite time, guaranteeing a positive lower bound on inter-execution times. Thus the occurrence of Zeno behavior is excluded.

Next, we proceed with a comparison between the proposed triggering mechanism and those introduced in \cite{xing2023dynamic,tang2022event}. 

On the one hand, unlike the dynamic event-triggered control presented in \cite{xing2023dynamic} for deterministic strict-feedback nonlinear systems, this work focuses on the more intricate stochastic non-triangular interconnected nonlinear systems. This scenario is inherently more challenging than the one considered in \cite{xing2023dynamic} due to two main aspects.
First, for stochastic nonlinear systems, the direct application of the existing event-triggering mechanisms utilized in deterministic nonlinear systems \cite{xing2023dynamic} may result in the occurrence of Zeno behavior. This is because the states of stochastic nonlinear systems can exceed any bound in an arbitrarily small amount of time, potentially resulting in the satisfaction of the triggered conditions within an arbitrarily small interval. Consequently, ensuring a minimum positive inter-execution time is non-trivial and demands an innovative approach to effectively exclude the occurrence of Zeno behavior.
Second, the nonlinear coupling interactions among subsystems exist in the entire interconnected system and the interactions to one subsystem usually depend on the states of all subsystems (i.e., do not adhere to the triangular structure of system states). Addressing the effects of nonlinear coupling interactions poses a significant challenge in the design and analysis of local controllers which only employ respective locally intermittent output signal for feedback, especially when these interactions lack a triangular structure.

On the other hand, the proposed triggering mechanism enhances practicality compared to the existing time-regulation-based method \cite{tang2022event}.
According to (\ref{eq:A62}), with a fixed clock variable value $\Pi_{y_i}$, a larger triggering error $\varpi_{y_i}$ induces a higher rate of change in $\dot{\Pi}_{y_i}$. This accelerates the reduction of ${\Pi}_{y_i}$ from $\bar{\Pi}_{y_i}$ to $\underline{\Pi}_{y_i}$, thereby contributing to the fulfillment of the triggering condition (\ref{eq:55}). Typically, the triggering error $\varpi_{y_i}$ is relatively large during the initial system startup phase and decreases as the system approaches stability. This, in conjunction with the proposed triggering mechanism, results in more frequent triggers during the initial startup phase and fewer triggers as the system stabilizes. This triggering behavior aligns more closely with practical engineering requirements. Therefore, the proposed triggering mechanism is more practical compared to the time-regulation-based method  \cite{tang2022event}, where the additional clock variable is independent of the triggering error, allowing for the exact computation of the minimum inter-execution times.}}


\subsection{Decentralized Dynamic Event-triggered Control Scheme}
In the context of dynamic event-triggered communication, we present a {\color{black}{decentralized output-feedback control scheme}} using only the locally intermittent output signal:
\begin{align}
v_{i}=&-\kappa_i
\left(b_{i,n_i}\bar{y}_i
+\frac{1}{L_{i,1}L_{i,2}}b_{i,n_i-1}
\hat{x}_{i,2}
+\cdots\right.& \nonumber\\
&\left. +\frac{b_{i,2}}{(L_{i,1}L_{i,2})^{n_i-2}}
\hat{x}_{i,n_i-1}
+\frac{b_{i,1}}{(L_{i,1}L_{i,2})^{n_i-1}}
\hat{x}_{i,n_i}  \right),
&\label{eq:69}
\end{align}
where $\bar{y}_i$ is the intermittent output signal (as illustrated in (\ref{eq:M55})), $\kappa_i={\left(L_{i,1}L_{i,2}\right)^{n_i}}$, and $b_{i,k}$ is the positive coefficient of the Hurwitz polynomial $h_2(s) = s^{n_i} + b_{i,1}s^{n_i-1} + \cdots + b_{i,n_i-1}s + b_{i,n_i}$, $k = 1,\cdots,n_i$.
{\color{black}{The block diagram of the proposed decentralized dynamic event-triggered control strategy is depicted in Fig. 1.}}
\begin{figure}
\begin{center}
\includegraphics[width=0.45\textwidth,height=42mm]{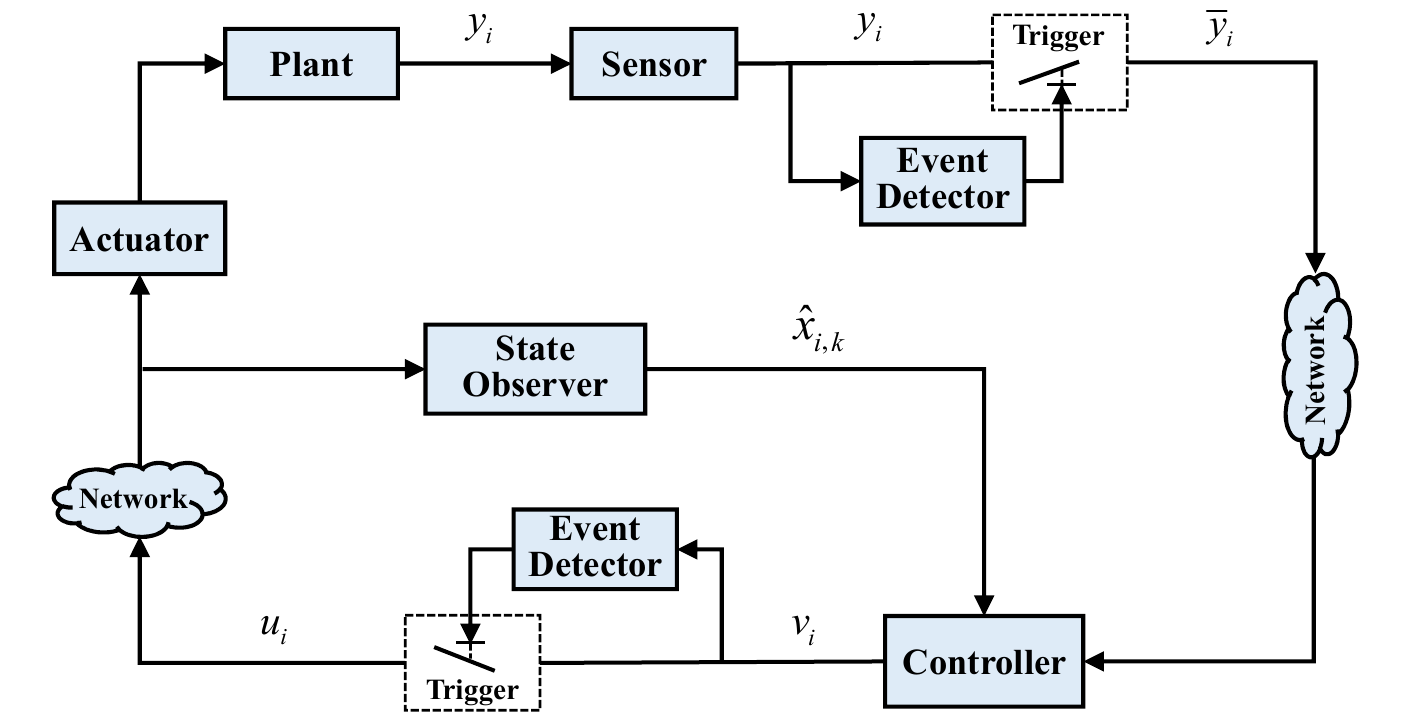}
\caption{A block diagram of the proposed decentralized control strategy with output and controller triggering.}
\end{center}
\end{figure}

Now, we proceed to state our main result in the following theorem.
\begin{theorem}\label{theorem2}
Consider the stochastic non-triangular nonlinear interconnected system (\ref{eq:1}) under Assumptions~\ref{assumption1}-\ref{assumption3}. If the state observer (\ref{eq:11}), {\color{black}{the modified clock-based dynamic triggering mechanism (\ref{eq:M55})-(\ref{eq:B62}),}} and the decentralized output-feedback controller  (\ref{eq:69}) are employed, the following conclusions hold:
\begin{itemize}
\item [i)]{The global asymptotic stability of the closed-loop system in probability is guaranteed.}
\item [ii)]{The states and the output of each local subsystem are ensured to converge to the equilibrium at the origin in probability.}
\item [iii)]{The occurrence of Zeno behavior is eliminated.}
\end{itemize}
\end{theorem}

{\emph{Proof}}. The proof consists of two parts: stability analysis and determination of the constant gains.

Part I. Stability analysis. Consider the Lyapunov function candidate
\begin{align}
V= V_e + V_z + V_{\Xi},&\label{eq:68}
\end{align}
where {\color{black}{$V_e=\sum_{i=1}^{N}V_{e_i}$, $V_z=\sum_{i=1}^{N}V_{z_i}$, and
$V_{\Xi}=\sum_{i=1}^{N}{\Xi}_{y_i}+\sum_{i=1}^{N}
{\Xi}_{u_i}$, with $V_{e_i}={\color{black}{\Pi_i}}e_i^{T}P_ie_i$,  $V_{z_i}={\color{black}{\Pi_i}}z_i^{T}Q_iz_i$, and $\Pi_i=\Pi_{y_i}+\Pi_{u_i}$.
From (\ref{eq:1}), (\ref{eq:11}), (\ref{eq:14}), and (\ref{eq:15}), and applying the ${\rm{It\hat{o}}}$'s differentiation rule, $\mathcal{L}V_{e_i}$ is derived as
\begin{align}
\mathcal{L} V_{e_i}=2{\Pi_i} e_i^TP_i(L_{i,1}
{A_e}{e_i}+L_{i,1}&G_ex_{i,1}+F_e)&\nonumber\\
+&\dot{\Pi}_ie_i^TP_ie_i+{\Pi}_iP_i\left\|
K_e\right\|^2.&\label{eq:C20}
\end{align}
where the matrices $A_e$, $G_e$, $F_{e}$, and $K_{e}$ are defined before. By employing}} Young's inequality, it has
\begin{align}
2{\color{black}{\Pi_i}}L_{i,1}e_i^TP_i{A_e}{e_i}\le& -{\color{black}{\Pi_i}}L_{i,1}\|e_i\|^2,
&\label{eq:C21}\\
2\Pi_iL_{i,1}e_i^TP_i{G_e}x_{i,1}
{\rm{\le}}&\frac{1}{2}\Pi_iL_{i,1}\left\|e_i\right\|^2
{\rm{+}}2\Pi_iL_{i,1}{r_e^2}\|P_i\|^2\|z_{i}\|^2,
&\label{eq:C22}\\
2{\color{black}{\Pi_i}}e_i^TP_i{F_e}\le& \frac{1}{2}{\color{black}{\Pi_i}}\|P_i\|^2\|e_i\|^2
{\rm{+}}\sum_{j=1}^{N}\sum_{k=1}^{n_i}\frac{2{\color{black}{\Pi_i}}\hbar_{ij,k}^2}{L_{i,1}
^{2(k{\rm{-}}1)}}\epsilon_{x_j}\|z_{j}\|^2&\nonumber\\
&+\sum_{j=1}^{N}\sum_{k=1}^{n_i}\frac{2{\color{black}{\Pi_i}}}{L_{i,1}^{2(k-1)}}\hbar_{ij,k}^2\sigma_{x_j}\|e_{j}\|^2,&\label{eq:C23}\\
{\color{black}{\Pi_i}}P_i\|K_e\|^2\le & \sum_{k=1}^{n_i}\frac{{\color{black}{\Pi_i}} }
{L_{i,1}^{2(k-1)}}k^2\ell_{i,k}^2\|P_i\|(\epsilon_{x_i}\|z_{i}\|^2
{\rm{+}}\sigma_{x_i}\|e_{i}\|^2),&\label{eq:C24}\\
{\color{black}{\dot{\Pi}_ie_i^TP_ie_i}}\le& \delta_{i}\Pi_{i}^2\|P_i\|\|e_{i}\|^2,
&\label{eq:C25}
\end{align}
where $\delta_{i}=\max\{\delta_{u_i},\delta_{y_i}\}$ is a positive constant. According to the definition of $V_e$ and using (\ref{eq:C20})-(\ref{eq:C25}), we obtain
\begin{align}
\mathcal{L}V_{e}\le -\sum_{j=1}^{N}{\color{black}{\Pi_j}}
\sigma_{e_j}^{*}\left\|e_{j}\right\|^2+\sum_{j=1}^{N}
{\color{black}{\Pi_j}}\sigma_{z_j}^{*}\left\|z_{j}\right\|^2,
&\label{eq:C27}
\end{align}
where $\sigma_{e_j}^{*}=\frac{1}{2}L_{j,1}
-{\color{black}{\delta_j\bar{\Pi}_{j}\left\|P_j\right\|}}
-2\sum_{i=1}^{N}{\color{black}{\sum_{k=1}^{n_i}
\frac{1}{L_{j,1}^{2(k-1)}}\hbar_{ij,k}^2\sigma_{x_j}}}
-\frac{1}{2}\left\|P_j\right\|^2-{\color{black}{\sum_{k=1}^{n_i} }}\frac{k^2}{L_{j,1}^{2(k-1)}}\ell_{j,k}^2
\sigma_{x_j}\left\|P_j\right\|$ and $\sigma_{z_j}^{*}=2L_{j,1}{r_e^2}\left\|P_j\right\|^2
+\sum_{i=1}^{N}\sum_{k=1}^{n_i}\frac{2}{L_{j,1}^{2(k-1)}}\hbar_{ij,k}^2\epsilon_{x_j}
+{\color{black}{\sum_{k=1}^{n_i}}} \frac{k^2}{L_{j,1}^{2(k-1)}}\ell_{j,k}^2 \epsilon_{x_j}\left\|P_j\right\|$, {\color{black}{with $\bar{\Pi}_{j}=\bar{\Pi}_{y_j}+\bar{\Pi}_{u_j}$.}}

From (\ref{eq:1}), (\ref{eq:11}),  (\ref{eq:12}), (\ref{eq:13}), (\ref{eq:34}), and (\ref{eq:69}), and using ${\rm{It\hat{o}}}$'s differentiation rule, there holds
\begin{align}
\mathcal{L} V_{z_i}=&{\color{black}{\dot{\Pi}_i}}z_i^TQ_iz_i+
2{\color{black}{\Pi_i}}z_i^TQ_i\left(L_{i,1}L_{i,2}
{A_z}{z_i}+\frac{L_{i,1}}{L_{i,2}}G_z(e_{i,1}
{\rm{-}}z_{i,1})\right.& \nonumber\\
&\left. +F_z+L_{i,1}L_{i,2}{R}_z{b_{i,n_i}}
\left(1-\theta_i\right)
z_{i,1}+L_{i,1}D_ze_{i,2}\right.& \nonumber\\
&\left.+\frac{1}{\left(L_{i,1}L_{i,2}\right)
^{n_i-1}}(K_u+ \Gamma_u ) \right) +{\color{black}{\Pi_i}}Q_i\left\|K_z\right\|^2,
&\label{eq:M82}
\end{align}
where the matrices $A_z$, $R_z$, $G_z$, $D_z$, $F_z$, and $K_z$ are defined before,
$K_{u}=\left[0,\,\cdots,\,0,\,u_i-v_i\right]^{T}$, and $\Gamma_{u}=\left[0,\,\cdots,\,0,\,\kappa_i b_{i,n_i}({y}_i-\bar{y}_i)\right]^{T}$.
{\color{black}{By employing the Young's inequality, the following inequalities hold true}}
\begin{align}
{\color{black}{\dot{\Pi}_i z_i^TQ_iz_i\le }}&{\color{black}{\,\delta_i\Pi_i^2\|Q_i\|\|z_{i}\|^2}}
&\label{eq:C47}\\
2{\color{black}{\Pi_i}}L_{i,1}L_{i,2}z_i^TQ_i{A_z}{z_i}
\le&-{\color{black}{\Pi_i}}L_{i,1}L_{i,2}\|z_i\|^2,&\label{eq:C40}\\
2{\color{black}{\Pi_i}}z_i^TQ_iL_{i,1}L_{i,2}{R}_z{b_{i,n_i}}&
(1-\theta_i)z_{i,1} &\nonumber\\
\le & 2{\color{black}{\Pi_i}}L_{i,1}L_{i,2}{b_{i,n_i}}\|Q_i\|
|1{\rm{-}}\theta_i|\|z_{i}\|^2,&\label{eq:C41}\\
2{\color{black}{\Pi_i}}\frac{L_{i,1}}{L_{i,2}}z_i^TQ_iG_z e_{i,1}
\le &\frac{1}{2}{\color{black}{\Pi_i}}\|Q_i\|^2\|e_i\|^2{\rm{+}}2{\color{black}{\Pi_i}}
\frac{L_{i,1}^2}{L_{i,2}^2}r_z^2\|z_i\|^2,&\label{eq:C42}\\
-2{\color{black}{\Pi_i}}\frac{L_{i,1}}{L_{i,2}}z_i^TQ_iG_z z_{i,1}
\le & 2{\color{black}{\Pi_i}}\frac{L_{i,1}}{L_{i,2}}r_z \left\|Q_i\right\|\left\|z_i\right\|^2,
&\label{eq:C43}\\
2{\color{black}{\Pi_i}}L_{i,1}z_i^TQ_iD_ze_{i,2}\le &\frac{L_{i,1}}{2}{\color{black}{\Pi_i}}
{\color{black}{\|Q_i\|^2}}\|e_i\|^2{\rm{+}}2
{\color{black}{\Pi_i}}L_{i,1} \|z_i\|^2,&\label{eq:C44} \\
2{\color{black}{\Pi_i}}z_i^TQ_i{F_z}\le&
\sum_{j=1}^{N}\frac{1}{2}{\color{black}{\Pi_i}}
\hbar_{ij,1}^2\left(\epsilon_{x_j}\left\|z_{j}\right\|^2
+\sigma_{x_j}\|e_{j}\|^2\right)&\nonumber\\
&+2{\color{black}{\Pi_i}}\|Q_i\|^2\|z_{i}\|^2,&\label{eq:C45}\\
{\color{black}{\Pi_i}}Q_i\left\|K_z\right\|^2
\le& {\color{black}{\Pi_i}}\ell_{i,1}^2\|Q_i\|\|z_{i}\|^2,&\label{eq:C46}\\
\frac{2{\color{black}{\Pi_i}}z_i^TQ_i}{\left(L_{i,1}L_{i,2}\right)^{n_i-1}}K_u\le & {\frac{{\color{black}{\Pi_i^2}}\left\|Q_i\right\|^2}{\left( L_{i,1}L_{i,2}\right)^{2(n_i-1)}}}
\left\|z_i\right\|^2+ \varpi_{u_i}^2, &\label{eq:a87}\\
\frac{2{\color{black}{\Pi_i}} z_i^TQ_i}{(L_{i,1}L_{i,2})^{n_i-1}}\Gamma_u
\le&{\color{black}{\Pi_i^2}}\|Q_i\|^2L_{i,1}^{2}L_{i,2}^{2}b_{i,n_i}^2\|z_i\|^2
+\varpi_{y_i}^2.&\label{eq:87}
\end{align}
{\color{black}{By the definition of $V_z$ and using (\ref{eq:M82})-(\ref{eq:87}), it holds that
\begin{align}
\mathcal{L} V_{z}\le\sum_{j=1}^{N}\Pi_j\rho_{e_j}^{*}\|e_j\|^2{\rm{-}}\sum_{j=1}^{N}
{\color{black}{\Pi_j}}{\color{black}{\rho_{z_j}^*}} \|z_j\|^2{\rm{+}}\sum_{j=1}^{N}\varpi_{u_j}^2+\sum_{j=1}^{N}\varpi_{y_j}^2,
&\label{eq:89}
\end{align}
where}} $\rho_{e_j}^*=\frac{1}{2}L_{j,1}
{\color{black}{\|Q_j\|^2}}+{\color{black}{\frac{1}{2} \|Q_j\|^2}}+\sum_{i=1}^{N}\frac{1}{2} \hbar_{ij,1}^2\sigma_{x_j}$ and $\rho_{z_j}^*=
L_{j,1}L_{j,2}{\color{black}{-\delta_j\bar{\Pi}_j\|Q_j
\|}}-2\frac{L_{j,1}^2}{L_{j,2}^2}r_z^2
-\frac{2L_{j,1}}{L_{j,2}}r_z\|Q_j\|-2{\color{black}{
\|Q_j\|^2}}-\ell_{j,1}^2\|Q_j\|{\rm{-}}
{\color{black}{2L_{j,1} }}{\rm{-}}\sum_{i=1}^{N}\frac{1}{2} \hbar_{ij,1}^2\epsilon_{x_j}{\rm{-}}2L_{j,1}L_{j,2}
{b_{j,n_j}}\|Q_j\|\left|1-\theta_j\right|
-{\color{black}{{\frac{1}{(L_{j,1}L_{j,2})
^{2(n_j-1)}}}\Pi_j\|Q_j\|^2-\Pi_j\|Q_j\|^2L_{j,1}^2L_{j,2}^2
b_{j,n_j}^2}}$.
Synthesizing (\ref{eq:59}), (\ref{eq:62}), {\color{black}{(\ref{eq:A62}), (\ref{eq:B62}),}} (\ref{eq:68}), {\color{black}{(\ref{eq:C27}), and (\ref{eq:89}), gives rise to}}
\begin{align}
\mathcal{L} V\le & {\color{black}{-\sum_{j=1}^{N}{\color{black}{\Pi_j}}
(\sigma_{e_j}^{*}{\rm{-}}\rho_{e_j}^*)\|e_{j}\|^2 {\rm{-}}\sum_{j=1}^{N}{\color{black}{\Pi_j}}
(\rho_{z_j}^{*}{\rm{-}}\sigma_{z_j}^{*}) \|z_j\|^2}}{\rm{-}}\sum_{j=1}^{N}
\beta_{y_j}\Xi_{y_j}&\nonumber\\
&-\sum_{j=1}^{N}\beta_{u_j}\Xi_{u_j}+ \sum_{j=1}^{N}\left(1-\gamma_{y_j}\right)
\varpi_{y_j}^2+\sum_{j=1}^{N}
\left(1-\gamma_{u_j}\right)\varpi_{u_j}^2&\nonumber\\
&+\sum_{j=1}^{N}\gamma_{y_j}\alpha_{y_j}
\left\|z_j^{*}\right\|^2+\sum_{j=1}^{N}
\gamma_{u_j}\alpha_{u_j}\left\|z_j^{*}\right\|^2.
&\label{eq:91}
\end{align}
{\color{black}{According to $\varpi_{y_j}^2\le \alpha_{y_j}\|z_j^{*}\|^2+ {\frac{\Xi_{y_j}}{\rho_{y_j}}}$ in the internal $[t_{y,l}^{j}, t_{y,l+1}^{j})$,  $\varpi_{u_j}^2\le \alpha_{u_j}\|z_j^{*}\|^2+ {\frac{\Xi_{u_j}}{\rho_{u_j}}}$ in the internal $[t_{u,l}^{j}, t_{u,l+1}^{j})$, ${\rho_{y_j}}>\frac{1- \gamma_{y_j}}{\beta_{y_j}}$, ${\rho_{u_j}}>\frac{1- \gamma_{u_j}}{\beta_{u_j}}$, and $\|z_j^{*}\| \le \|z_j\|$, $\mathcal{L} V$ in (\ref{eq:91}) becomes
\begin{align}
\mathcal{L} V\le -\sum_{j=1}^{N}{\color{black}{\Pi}_j}\xi_j^{*}
\left\|e_{j}\right\|^2-\sum_{j=1}^{N}{\color{black}{\Pi}_j}\varsigma_j^{*} \left\|z_{j}\right\|^2,
&\label{eq:98}
\end{align}
where $\xi_j^{*}=\sigma_{e_j}^{*}-\rho_{e_j}^{*}$ and $\varsigma_j^{*}=\rho_{z_j}^{*}
-\sigma_{z_j}^{*}-\alpha_{y_j}-\alpha_{u_j}$}}.

Part II. Determination of the constant gains. Following a similar vein to the continuous case in Section IV, the design recipe for $L_{j,1}$ and $L_{j,2}$ can be given under the dynamic event-triggered communication, i.e.,
\begin{align}
L_{j,1}=&\max\left\{1,\frac{2\hat{c}_1^{*}}{1-{\color{black}{\left\|Q_j\right\|^2}}}
\right\},&\label{eq:R11}\\
L_{j,2}\in &\left(\frac{1}{2}\tilde{\kappa}_1-\hat{\kappa},\frac{1}{2}\tilde{\kappa}_1
+\hat{\kappa}\right) \cap \left(1,+\infty\right),&\label{eq:R19}
\end{align}
where $\hat{c}_1^{*}={\color{black}{\delta_j
\bar{\Pi}_{j}\left\|P_j\right\|}}
+\frac{1}{2}\|P_j\|^2+{\color{black}{2}}
\sum_{i=1}^{N}{\color{black}{\bar{\hbar}_{ij,k}^2}}
\sigma_{x_j}+k^2{\color{black}{\bar{\ell}_{j,k}^2}}
\sigma_{x_j}\|P_j\|+\frac{1}{2}\|Q_j\|^2
+\sum_{i=1}^{N}\frac{1}{2}
{\color{black}{\bar{\hbar}_{ij,1}^2}}
\sigma_{x_j}>0$ is a computable constant, ${\color{black}{\hat{\kappa}}}=
\left(\frac{1}{4}\tilde{\kappa}_1^2
-{\color{black}{\tilde{\kappa}_1}}
\tilde{\kappa}_2\right)^{\frac{1}{2}}$,
${\color{black}{\tilde{\kappa}_1= \frac{1}{a_5 L_{j,1}}}}$, ${\color{black}{\tilde{\kappa}_2}}
=\frac{1}{\Lambda_jL_{j,1}}{\color{black}{\hat{c}_2^{*} }}+a_1 L_{j,1}+a_2+a_3+a_4$,
$\hat{c}_2^{*}={\color{black}{\alpha_{y_j}
+\alpha_{u_j} }} + {\color{black}{\delta_j
\bar{\Pi}_j\|Q_j\|}}+ {\color{black}{\bar{\ell}
_{j,1}^2}}\|Q_j\|
+2\left\|Q_j\right\|^2+\sum_{i=1}^{N}\frac{1}{2} {\color{black}{\bar{\hbar}_{ij,1}^2}}\epsilon_{x_j}
+\sum_{i=1}^{N}{\color{black}{2}}\bar{\hbar}_{ij,k}^2\epsilon_{x_j}
+k^2\bar{\ell}_{j,k}^2\epsilon_{x_j}\|P_j\|$,
$a_1={\color{black}{\frac{2}{\Lambda_j} }} r_z^2$, $a_2={\color{black}{\frac{2}{\Lambda_j} }}({r_e^2}\|P_j\|^2+1)$, $a_3=\frac{ 2}{\Lambda_j}r_z\|Q_j\|$, $a_4={\color{black}{\frac{1}{\Lambda_j}\bar{\Pi}_j}}\|Q_j\|^2$, and $a_5={\color{black}{\frac{1}{\Lambda_j} \bar{\Pi}_j}}\|Q_j\|^2
b_{j,n_j}^2$.

{\color{black}{In what follows, we show the validity of the results presented in Theorem~\ref{theorem2}.

From (\ref{eq:98}) and utilizing the design recipe of constant gains provided in (\ref{eq:R11}) and (\ref{eq:R19}), we obtain $\mathcal{L} V \le -\xi_0 \left(\sum_{i=1}^{N} {\Pi}_i \left\|e_{i}\right\|^2+\sum_{i=1}^{N} {\Pi}_i \left\|z_{i}\right\|^2\right)$,
where $\xi_0=\min\{\xi_i^{*}, \varsigma_i^{*}\}$ is a positive constant. This indicates that $\mathcal{L}V$ becomes the same form as (3.19) in \cite{deng2001stabilization} (resp, (38) in \cite{xie2009adaptive}). Besides, according to
Assumptions~\ref{assumption1}-\ref{assumption2}, for the stochastic nonlinear system (\ref{eq:1}), there exist class ${K}_{\infty}$ functions $\alpha_1$ and $\alpha_2$, such that $ \alpha_1\left(|(\cdot)|\right) \leq V(\cdot) \leq \alpha_2\left(|(\cdot)|\right)$. From (\ref{eq:68}) and by applying Theorem 3.1 in  \cite{deng2001stabilization} (resp, Lemma 1 in \cite{xie2009adaptive}), we establish that the closed-loop system is globally asymptotically stable in probability. Consequently, according to the definition of $V$ in (\ref{eq:68}), both $e_{i,k}$ and $z_{i,k}\,(k=1,\cdots,n_i)$ converge to the equilibrium at the origin in probability. Furthermore, from  (\ref{eq:33}) and (\ref{eq:34}), and considering that $y_i=\theta_ix_{i,1}$ (with $1-\bar{\theta}_{i}\le \theta_i\le 1+\bar{\theta}_{i}$), it follows that the states $x_{i,k} \, (k=1,\cdots,n_i)$ and the output $y_i$ of each local subsystem converge to the equilibrium at the origin in probability. Therefore, the results i)-ii) in Theorem 2 are established.}}

{\color{black}{Next, we establish the validity of result iii) in Theorem 2. According to the proposed modified clock-based dynamic triggering mechanism, an event is triggered only when the conditions of both parts in (\ref{eq:55}) (resp, (\ref{eq:58})) are concurrently satisfied. If either condition fails to meet, the event will not be triggered. Since the second part of the triggering condition (\ref{eq:55}) (resp, (\ref{eq:58})) guarantees a positive lower bound on inter-execution times (as detailed in Section V-B), there exist positive constants $T_{y}^{*}$ and $T_{u}^{*}$ such that $t_{y,l+1}^{i}-t_{y,l}^{i}> T_{y}^{*}$ and $t_{u,l+1}^{i}-t_{u,l}^{i}>T_{u}^{*}$. Hence, the occurrence of Zeno behavior is effectively prevented.}} $\hfill\blacksquare$

\begin{remark}\label{remark8}
While ensuring system stability, it is of paramount importance to enhance system performance \cite{lv2022prescribed,lv2023distributed}.
The introduction of the event-triggering mechanism inherently induces triggering errors, i.e., the errors between the triggering signals and the continuous signals. These errors bear the potential to exert a substantial influence on the performance of the system, or even damage the system stability. Notably, existing event-triggered control results  \cite{wang2021adaptive,sun2023decentralized,sun_2022_distributed}  can only guarantee bounded stabilization/tracking errors.
Achieving zero stabilization/tracking errors within the event-triggered control framework is a critical yet challenging objective for meeting the accuracy requirements of the overall system. The underlying problem becomes even more complicated when the system involves severe uncertainties, including stochastic disturbances, non-triangular structural uncertainties, and unknown time-varying sensor sensitivity. For this purpose, this study introduces an innovative decentralized output-feedback control strategy with a modified clock-based dynamic triggering mechanism, ensuring that the closed-loop system is globally asymptotically stable in probability. 
\end{remark}


\section{Simulation Verification}
To examine the effectiveness of the proposed approach, we consider the following numerical example:
\begin{align}
d{x}_{i,1}=\,&\left(x_{i,2}+\sum_{j=1}^{2} f_{i j, 1}\right)dt+{\color{black}{\phi_{i,1}^{T}}} (x_{i,1})dw_i,&\nonumber\\
d{x}_{i,2}=\,&\left(u_{i}+\sum_{j=1}^{2} f_{i j, 2}\right){\color{black}{dt}}
+{\color{black}{\phi_{i,2}^{T}}}(\check{x}_{i,2})dw_i,&\nonumber\\
y_{i}=\,&\theta_i(t)x_{i,1},&\label{eq:73}
\end{align}
where the sensor sensitivity $\theta_i(t)=1+0.25\left|\sin(10t)\right|$, the nonlinearities  ${\color{black}{\phi_{i,1}}}(x_{i,1})=0.5x_{i,1}\sin(0.2x_{i,1})$ and ${\color{black}{\phi_{i,2}}}(\check{x}_{i,2})=
0.25x_{i,1}^2x_{i,2}^2\cos(x_{i,2})$.
We set the initial states $x_{i,1}(0)=x_{i,2}(0)=0.1$,
$\hat{x}_{i,1}(0)=\hat{x}_{i,2}(0)=0.2$,
$\Xi_{y_i}(0)=\Xi_{u_i}(0)=0$,
the design parameters $\beta_{y_i}=\beta_{u_i}=1$, {\color{black}{$\rho_{y_i}=\rho_{u_i}=1$, $\alpha_{y_i}=\alpha_{u_i}=0.05$, $\gamma_{y_i}=\gamma_{u_i}=0.05$,}}
{\color{black}{$a_{i,1}=8$, $a_{i,2}=1$, $L_{i,1}=4$, $L_{i,2}=3$, $b_{i,1}=20$, $b_{i,2}=2$,}} $\bar{\Pi}_{y_i}=\bar{\Pi}_{u_i}=0.5$,
$\underline{\Pi}_{y_i}=\underline{\Pi}_{u_i}=0.3$, {\color{black}{$\delta_{y_i}=\delta_{u_i}=15$}}, $i=1,2$, the nonlinear coupling interaction terms $f_{11,1}=0.1\sin (u_1u_2)\Theta_1$,
$f_{12,1}=0.15 \Theta_2$,
$f_{11,2}=0.1 \Theta_1$,
$f_{12,2}=0.15 \sin\left(\Theta_2\right)$,
$f_{21,1}=f_{21,2}=0.15 \Theta_1$,
and $f_{22,2}=0.1 \ln \left(1+\Theta_2\right)$, with $\Theta_1=({x_{1,1}^{2}+x_{1,2}^{2}})^{1/2}$ and $\Theta_2=({x_{2,1}^{2}+x_{2,2}^{2}})^{1/2}$.
{\color{black}{To verify the effectiveness of the proposed triggering mechanism, we provide a comparison between the proposed method and the existing time-regulation-based method \cite{tang2022event}. We set the clock variable dynamics as: 1) The proposed method:  $\dot{\Pi}_{y_i}= -\delta_{y_i}  \left(\varpi_{y_i}^2+ \Pi_{y_i}^2\right)$ and $\dot{\Pi}_{u_i}=-\delta_{u_i}  \left(\ \varpi_{u_i}^2+ \Pi_{u_i}^2\right)$; and 2) The time-regulation-based method \cite{tang2022event}: $\dot{\Pi}_{y_i}= -\delta_{y_i} \Pi_{y_i}^2$ and $\dot{\Pi}_{u_i}=-\delta_{u_i}  \Pi_{u_i}^2$. All other design parameters remained the same for both methods to ensure a fair comparison.}}

The results of the proposed method are depicted in Fig. 2.  Fig. 2 (a) illustrates the evolution of states $\hat{x}_{i,k}$ and state estimates $\hat{x}_{i,k}$, from which we observe that $\lim_{t\to +\infty}\hat{x}_{i,k}=\lim_{t\to +\infty}x_{i,k}=0$. Fig. 2 (b) shows that the output $y_i$ converges to the equilibrium at the origin in probability. Fig. 2 (c) demonstrates the boundedness of the control input $u_i$. The dynamic auxiliary variables $\Xi_{y_i}$ and $\Xi_{u_i}$ are displayed in Fig. 2 (d).
{\color{black}{The comparison results are depicted in Fig. 3. Figs. 3 (a)-(b) showcase the results obtained with the proposed method. In Fig. 3 (a), the triggering times for $u_i$ and $y_i$ are depicted. Fig. 3 (b) shows the clock variables $\Pi_{y_i}$ and $\Pi_{u_i}$. In Figs. 3 (c)-(d), the results for the time-regulation-based method \cite{tang2022event} are presented. Fig. 3 (c) gives the triggering times for $u_i$ and $y_i$, and Fig. 3 (d) depicts  the clock variables $\Pi_{y_i}$ and $\Pi_{u_i}$.
From which it can be observed that both methods are able to effectively avoid the occurrence of Zeno behavior. Notably, the proposed method, which incorporates the triggering error terms $\varpi_{y_i}^2$ and $\varpi_{u_i}^2$ into the clock variable dynamics ${\Pi}_{y_i}$ and ${\Pi}_{u_i}$, respectively, leads to more frequent triggers during the initial system startup phase and fewer triggers as the system approaches stability. This triggering behavior aligns more closely with practical engineering requirements, thereby enhancing the practicality of the proposed method compared to the time-regulation-based method \cite{tang2022event}.}}


\begin{figure}[htbp]
\centering{
\subfigure[]
{\includegraphics[width=1.6in]{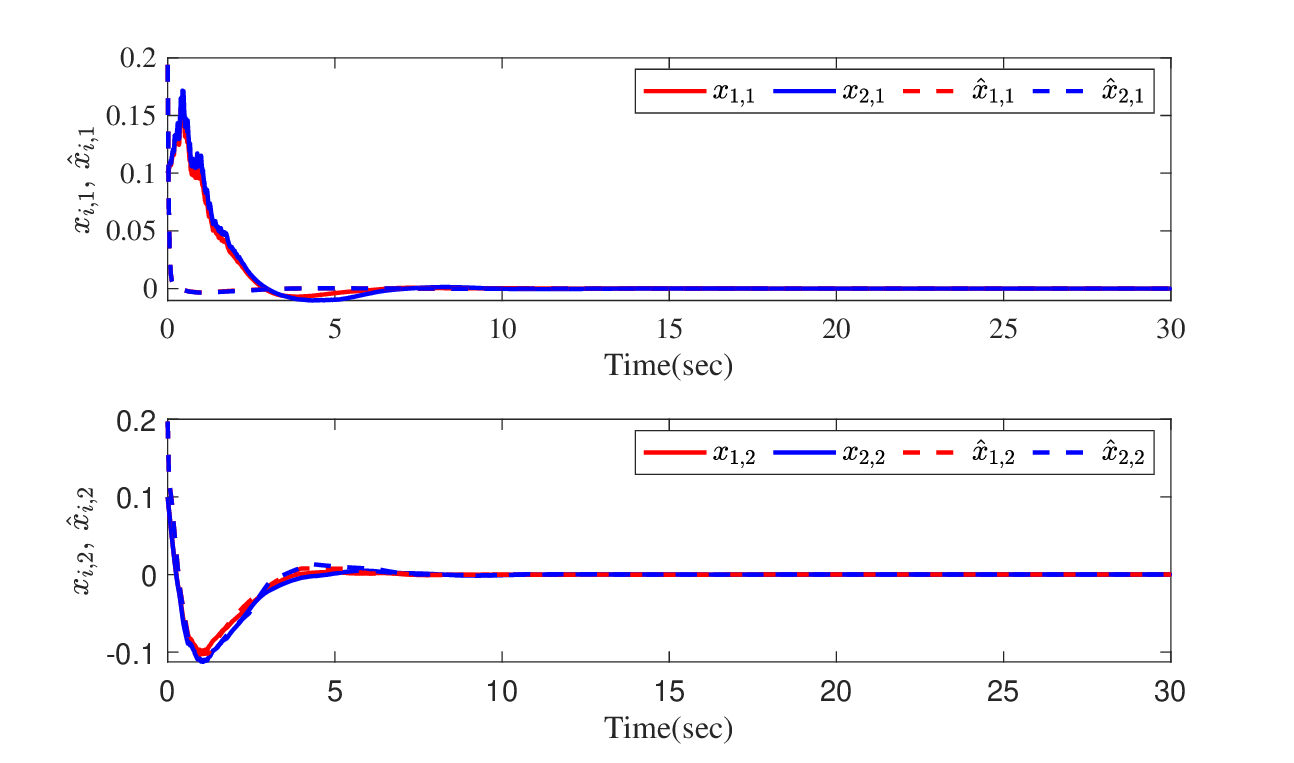}}
\subfigure[]
{\includegraphics[width=1.6in]{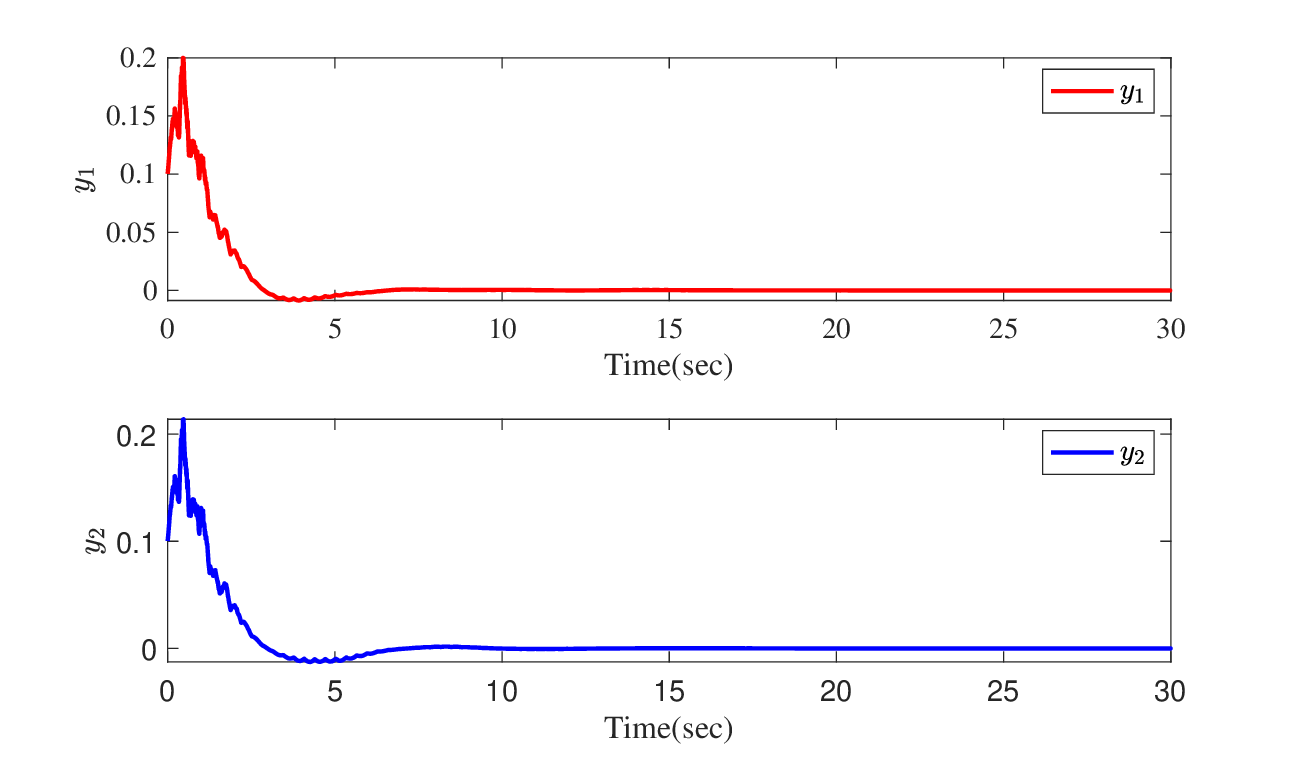}}
\subfigure[]
{\includegraphics[width=1.6in]{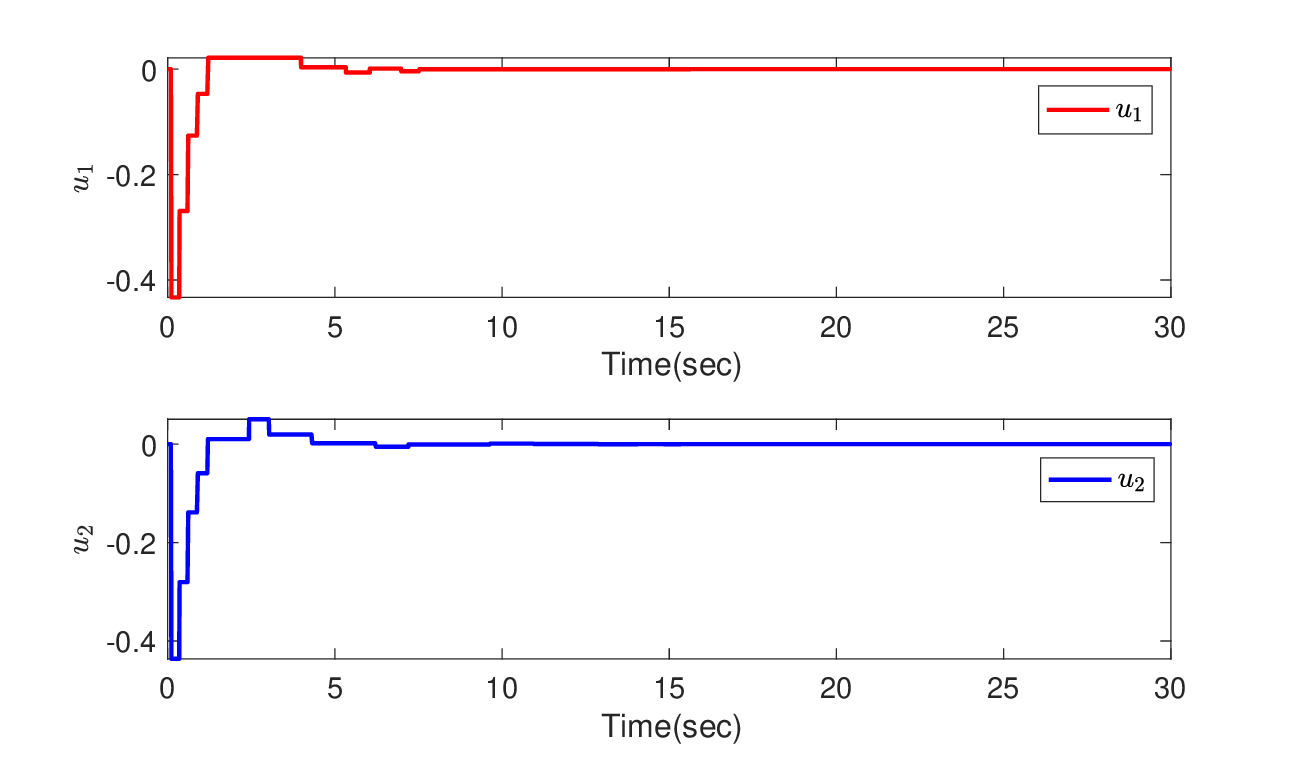}}
\subfigure[]
{\includegraphics[width=1.6in]{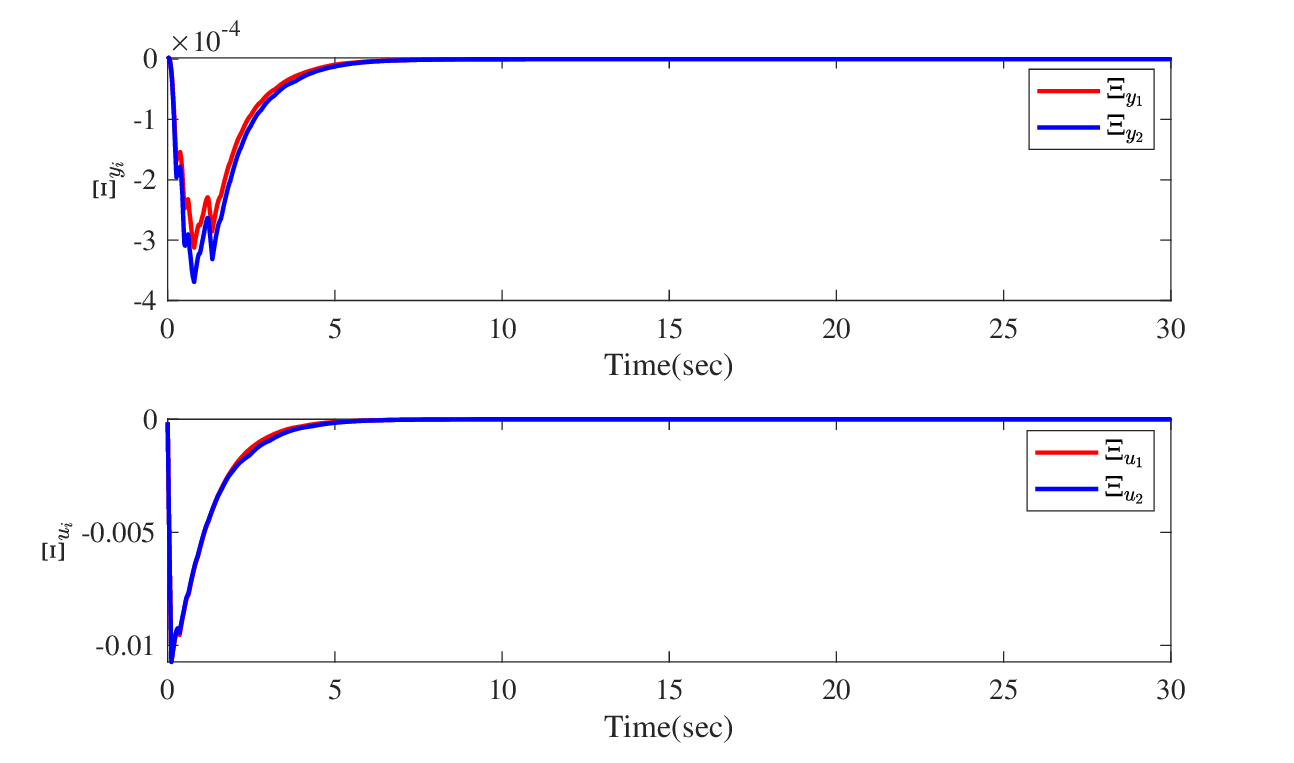}}}
\caption{{\color{black}{The results of the proposed method. (a) The states $x_{i,k}$ and state estimates $\hat{x}_{i,k}$. (b) The output $y_{i}$. (c) The control input $u_i$. (d) The dynamic auxiliary variables $\Xi_{y_i}$ and $\Xi_{u_i}$.}}}
\label{simulation_results}
\end{figure}
\begin{figure}[htbp]
	\centering{
		\subfigure[]
		{\includegraphics[width=1.6in]{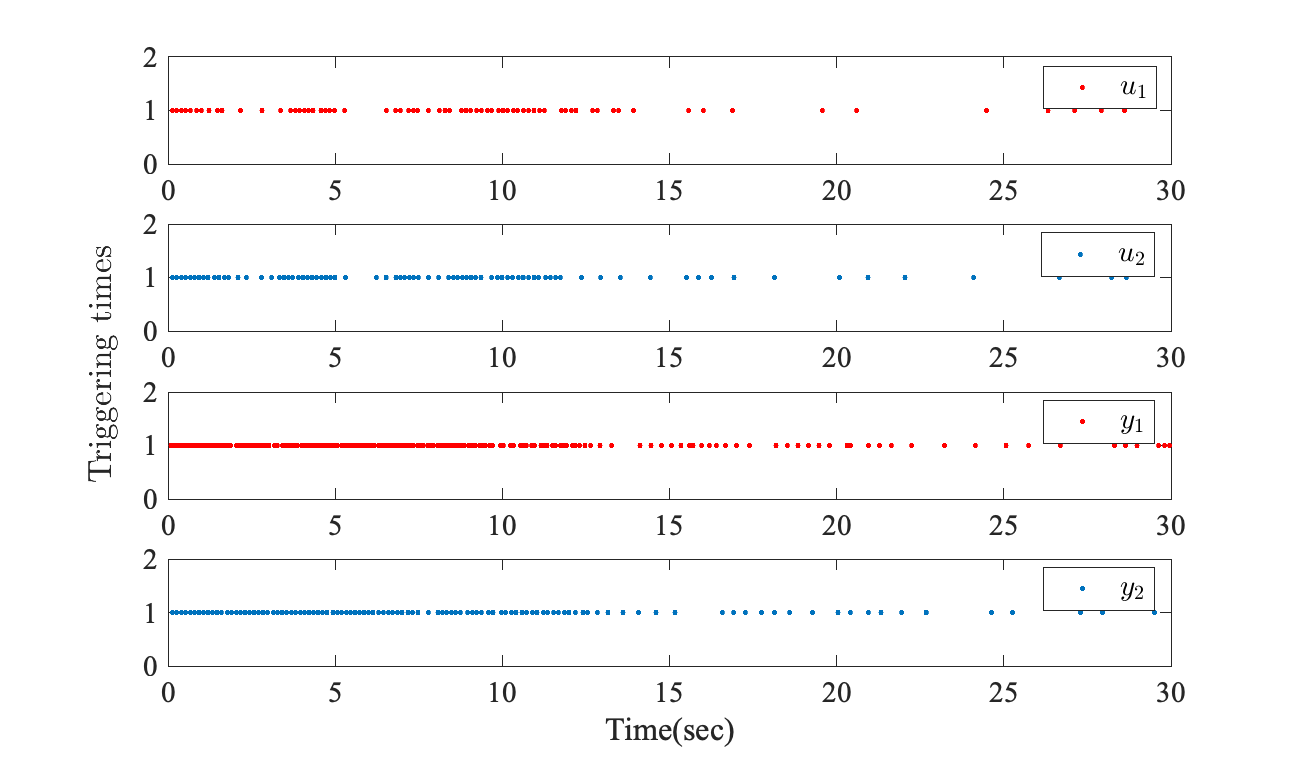}}
		\subfigure[]
		{\includegraphics[width=1.6in]{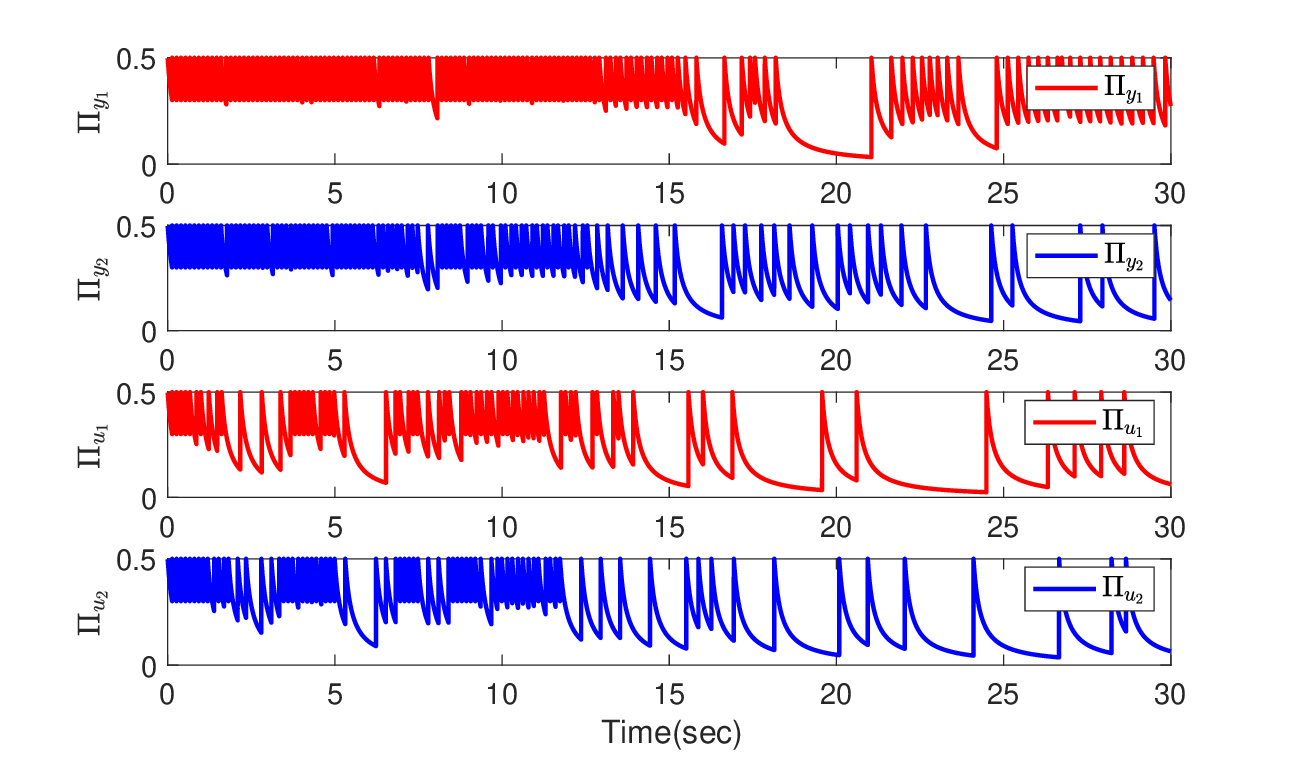}}
		\subfigure[]
		{\includegraphics[width=1.6in]{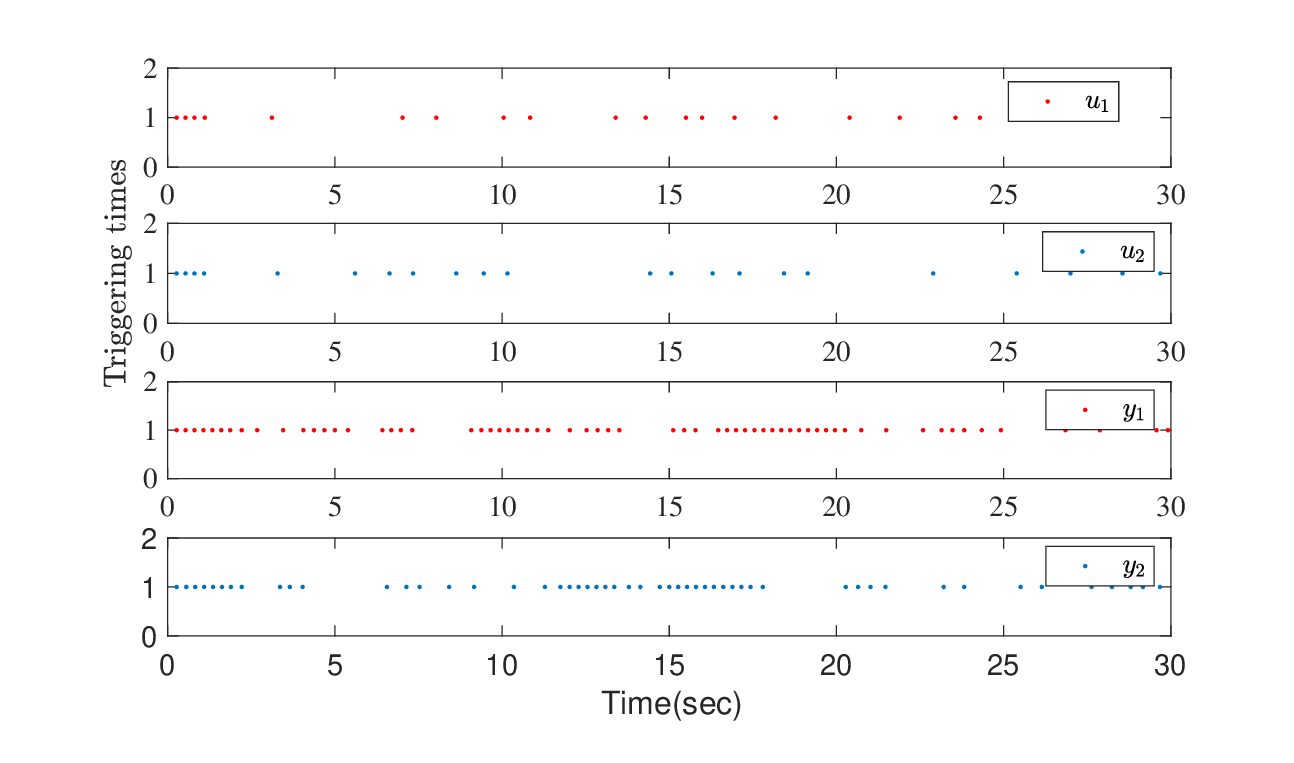}}
		\subfigure[]
		{\includegraphics[width=1.6in]{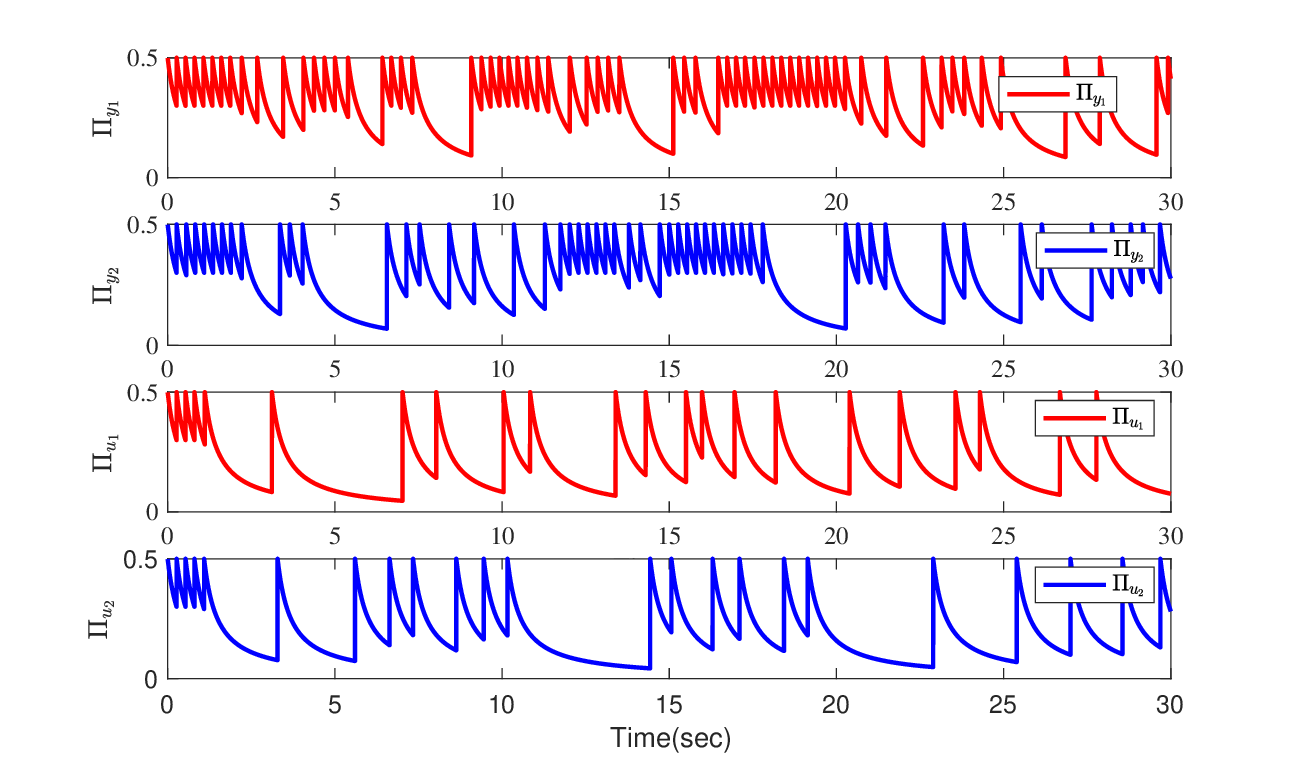}}}
	\caption{{\color{black}{The comparison results. (a) The triggering times of $u_i$ and $y_i$ under the proposed method. (b) The clock variables $\Pi_{y_i}$ and $\Pi_{u_i}$ under the proposed method. (c) The triggering times of $u_i$ and $y_i$ under the time-regulation-based method \cite{tang2022event} . (d) The clock variables $\Pi_{y_i}$ and $\Pi_{u_i}$ under the time-regulation-based method \cite{tang2022event} .}}}
	\label{simulation_results}
\end{figure}


\section{Conclusion}
This paper has introduced a {\color{black}{decentralized output-feedback control scheme}} for stochastic non-triangular nonlinear interconnected systems with event-triggered communication, where each local subsystem exchanges information with its neighbors, and only the locally intermittent output signal is utilized. Through the utilization of a novel coordinate transformation, we present a decentralized dynamic {\color{black}{event-triggered output-feedback control strategy,}} which involves a state observer and a decentralized {\color{black}{output-feedback controller.}} The analysis demonstrates that the global asymptotic stability of the closed-loop system in probability is guaranteed, with {\color{black}{the states and}} the output of each local subsystem converging to the equilibrium at the origin in probability. {\color{black}{Besides, the existence of a minimal dwell-time between triggering instants is guaranteed. An intriguing avenue for future research involves extending such results to achieve leader-following consensus for stochastic multi-agent systems.}}

\begin{appendices}
\section{}
\emph{Proof of Lemma~\ref{lemma3}}.
From (\ref{eq:12}), (\ref{eq:13}), and (\ref{eq:14}), it is seen that
\begin{align}
x_{i,1}=&z_{i,1},&\nonumber\\
x_{i,k}=& L_{i,1}^{k-1}e_{i,k}+{\left(L_{i,1}
L_{i,2}\right)^{k-1}}z_{i,k},k=2,\cdots,n_i. &\label{eq:a1}
\end{align}
By the definition of $x_i$ and using Young's inequality, there holds
\begin{align}
\left\|x_i\right\|^2\le& z_{i,1}^2+ \left(L_{i,1}e_{i,2}+L_{i,1}L_{i,2}z_{i,2}\right)^2
+\cdots&\nonumber\\
&+\left(L_{i,1}^{n_i-1}e_{i,n_i}+{\left(L_{i,1}
L_{i,2}\right)^{n_i-1}}z_{i,n_i}\right)^2,&\nonumber\\
\le& z_{i,1}^2+  2L_{i,1}^2e_{i,2}^2
+2(L_{i,1}L_{i,2})^2z_{i,2}^2+\cdots&\nonumber\\
&+2L_{i,1}^{2(n_i-1)}e_{i,n_i}^2+2{\left(L_{i,1}
L_{i,2}\right)^{2(n_i-1)}}z_{i,n_i}^2.&\label{eq:a2}
\end{align}
Since $\left|z_{i,k}\right|^2\le \left\|z_i\right\|^2$ and $\left|e_{i,k}\right|^2\le \left\|e_i\right\|^2$, $k=1,
\cdots,n_i$, it can be derived from (\ref{eq:a2}) that
\begin{align}
\left\|x_i\right\|^2 \leq  \epsilon_{x_i}\left\|z_{i}\right\|^2
+\sigma_{x_i}\left\|e_{i}\right\|^2, &\label{eq:a3}
\end{align}
where $\epsilon_{x_i}=1+2(L_{i,1}L_{i,2})^2
+\cdots+2{\left(L_{i,1}L_{i,2}\right)^{2(n_i-1)}}$ and $\sigma_{x_i}=2L_{i,1}^2+2L_{i,1}^{4}+\cdots
+2L_{i,1}^{2(n_i-1)}$.
$\hfill{\blacksquare}$

\section{}
\emph{Proof of Lemma~\ref{lemma4}}.
In light of Assumption~\ref{assumption1}, it holds that
\begin{align}
\sum_{j=1}^{N}\left|f_{i j, k}\left(x_{j}, u_{j},t\right)\right|
\le \sum_{j=1}^{N}\hbar_{i j, k}^2\left\|x_{j}\right\|^2. &\label{eq:b1}
\end{align}
According to the definition of $F_e$ and invoking Lemma~\ref{lemma3}, we have
\begin{align}
\left\|F_{e}\right\|=&\sum_{j=1}^{N}\left(f_{i j, 1}^2+\frac{1}{L_{i,1}^2}f_{i j, 2}^2
+\cdots+\frac{1}{L_{i,1}^{2(n_i-1)}}f_{i j, n_i}^2\right),&\nonumber\\
\le& \sum_{j=1}^{N}\sum_{k=1}^{n_i}\frac{1}{L_{i,1}^{2(k-1)}}\hbar_{ij,k}^2
\left(\epsilon_{x_j}\left\|z_{j}\right\|^2+\sigma_{x_j}\left\|e_{j}\right\|^2\right).
&\label{eq:b3}
\end{align}
In accordance with Lemma~\ref{lemma2} and Assumption~\ref{assumption2}, {\color{black}{one can obtain
\begin{align}
\left\|\phi_{i,1}\left(x_{i,1}\right)\right\|^2
\le&\ell_{i,1}^2\left|x_{i,1}\right|^2,&\nonumber\\
{\|\phi_{i,2}\left(\check{x}_{i,2}\right)\|^2}
\le &{2\ell_{i,2}^2}\left(\left|x_{i,1}\right|^2
+\left|x_{i,2}\right|^2 \right),&\nonumber\\
&  \vdots &\nonumber\\
{\|\phi_{i,n_i}\left({x}_{i}\right)\|^2}
\le& {n_i\ell_{i,n_i}^2} \left(|x_{i,1}|^2+\cdots+|x_{i,n_i}|^2\right).
&\label{eq:b6}
\end{align}
With}} (\ref{eq:b6}) in mind, we can derive from the definition of $K_e$ that
\begin{align}
\left\|K_e\right\|^2\le&
\ell_{i,1}^2\left|x_{i,1}\right|^2
+\frac{2}{L_{i,1}^2} \ell_{i,2}^2\left(\left|x_{i,1}
\right|^2+\left|x_{i,2}\right|^2 \right)&\nonumber\\
&+ \frac{n_i}{L_{i,1}^{2(n_i-1)}}\ell_{i,n_i}^2  \left(\left|x_{i,1}\right|^2+\cdots
+\left|x_{i,n_i}\right|^2 \right),&\nonumber\\
\le & \left(\ell_{i,1}^2+\frac{4}{L_{i,1}^2}
\ell_{i,2}^2{\rm{+}}\cdots {\rm{+}} \frac{n_i^2}{L_{i,1}
^{2(n_i-1)}}\ell_{i,n_i}^2 \right)
\left\|x_{i}\right\|^2. &\label{eq:b7}
\end{align}
This, {\color{black}{in conjunction with Lemma~\ref{lemma3}, establishes (\ref{eq:19}).}}
$\hfill{\blacksquare}$
\end{appendices}

\bibliographystyle{IEEEtran}
\bibliography{Stochastic_systems_20230903}
\end{document}